\documentclass[prb,twocolumn,longbibliography,showpacs]{revtex4-1}
\usepackage{graphicx}
\usepackage{latexsym}
\usepackage{graphicx}
\usepackage{times}
\usepackage{color}
\usepackage{amsmath}
\usepackage{dcolumn}
\usepackage{mathbbol}
\usepackage{latexsym,amsmath,amssymb,bm,euscript}
\usepackage{xr}
\usepackage{lipsum}
\usepackage[normalem]{ulem}

\usepackage{relsize}

\usepackage{hyperref}
\hypersetup{
    colorlinks,%
    citecolor=blue,%
    filecolor=blue,%
    linkcolor=blue,%
    urlcolor=blue
}

\DeclareMathOperator{\Tr}{Tr}

\begin{document}  
\title{Electron-photon interaction in a quantum point contact coupled to a microwave resonator}
\author{Udson C. Mendes}
\email{udsonmendes@gmail.com}
\author{Christophe Mora}
\email{mora@lpa.ens.fr}
\affiliation{Laboratoire Pierre Aigrain, \'Ecole normale sup\'erieure, PSL Research University, CNRS, Universit\'e Pierre et Marie Curie, Sorbonne
Universit\'es, Universit\'e Paris Diderot, Sorbonne Paris-Cit\'e, 24 rue Lhomond, 75231 Paris Cedex 05, France}

\date{\today} 

\begin{abstract}
We study a single-mode cavity weakly coupled to a voltage-biased quantum point contact. In a perturbative analysis, the lowest order predicts a thermal state for the cavity photons, driven by the emission noise of the conductor. The cavity is thus emptied as all transmission probabilities of the quantum point contact approach one or zero. Two-photon processes are identified at higher coupling, and pair absorption dominates over pair emission for all bias voltages. As a result, the number of cavity photons, the cavity damping rate and the second order coherence $g^{(2)}$ are all reduced and exhibit less bunching than the thermal state. These results are obtained with a Keldysh path integral formulation and reproduced with rate equations. They can be seen as a backaction of the cavity measuring the electronic noise. Extending the standard $P(E)$ theory to a steady-state situation, we compute the modified noise properties of the conductor and find quantitative agreement with the perturbative calculation.
\end{abstract}

\maketitle

\section{Introduction}

In the last years the research on circuit quantum electrodynamics (cQED) has been extended to mesoscopic devices, e.g., quantum dots (QDs) 
\cite{childress-lukin-pra2004,delbecq2011,Petersson2012,Schroer2012,BergenfeldtSET2012,frey-wallraff-prl2012,toida2013,basset2013,schiro2014,Liu2014}, 
tunnel junctions~\cite{gasse2013,soquet-simon-prb2013,Souquet2014,forgues2014,mendes-mora-njp2015,jin-schon-prb2015,qassemi-blais-prl2106,Parlavecchio2015,mora-portier-arxiv2105} 
or dc-biased Josephson junctions~\cite{leppaekangas-fogelstrom-prl2013,armour2013,gramich2013,chen2014,dambach2015}. The high sensitivity of the cavity 
field offers a powerful way to probe the electronic conductor in a non-invasive way~\cite{cottet2011,cottet2015,dmytruk-mora-simon-prb2015} thus acting as an excellent tool to investigate correlated electronic systems such as Majorana fermions~\cite{CottetMajos2013,QC32013,dmytruk-simon-prb2015} and Kondo physics 
\cite{delbecq2011,Deng-leHur-guo-arxiv2015}. Furthermore, many interesting phenomena due to the interplay between electrons and photons have been 
demonstrated in this hybrid device: quadrature squeezing \cite{gasse2013,forgues2014,mendes-mora-njp2015,qassemi-blais-prl2106,mora-portier-arxiv2105}, spin-photon 
coupling \cite{Petersson2012,viennot-kontos-science2015}, coupling between distant QDs \cite{delbecq2013,DT32013,lambert2013,deng2015}, and photon lasing 
\cite{jin-marthaler-schon-prb2011,gullans2015,lambert2015,liu2015}. These demonstrations open a huge avenue of possibilities in quantum information processing and quantum state engineering~\cite{souquet2016} with mesoscopic devices coupled to transmission lines. Also the photons emitted by a quantum conductor carry information on the dynamics of electrons~\cite{aguado-kouwenhoven-prl2000,basset2010,xu-belzig-prl2014}. They have been used to characterize the photonic side of the dynamical Coulomb blockade effect (DCB)~\cite{hofheinz-esteve-prl2011,altimiras2014,leppaekangas-fogelstrom-njp2014} in which photons are radiated by inelastic electron scattering.

Even with the rapid progress on cQED with mesoscopic devices the physics of a quantum point contact (QPC) coupled to a superconductor microwave 
resonator has been practically unexplored. The QPC is a coherent conductor formed by two metallic gates on the top of a two-dimensional electron gas. 
A voltage applied on the gates creates a one-dimension constriction connecting the two sides of the electron gas. The gate voltage controls both 
number of channels $n$ connecting the two metallic reservoirs and their transmission probabilities $T_{n}$. As microwave photons are created by 
the scattering of electrons tunneling from one reservoir to the other, the QPC is an excellent device to investigate photon statistics with 
controlled electron scattering. In the case of a coherent conductor QPC coupled to an open transmission line, it has been predicted that the 
photons emitted by the QPC present sub- or super-Poissonian statistics, depending on the transmission probability and source-drain voltage~\cite{beenakker-schomerus-prl2001,beenakker-schomerus-prl2004,fulga2010,lebedev-blatter-prb2010,hassler-otten-prb2015}. However, for a microwave cavity exchanging photons 
with the QPC, less is known about photon statistics. The cases of a tunnel junction~\cite{jin-schon-prb2015} and a quantum dot~\cite{BergenfeldtSET2012} 
have been investigated with a quantum master equation.

In this article, we discuss a dc-biased coherent QPC coupled to a microwave resonator formed with a transmission line cavity (TLC). 
A single mode is considered in the cavity, with frequency $f_0 = 2 \pi \omega_0$, such that our model equivalently describes a lumped LC circuit 
coupled to the QPC. When the dc-bias $V$ is smaller than the cavity frequency $\hbar \omega_0/e$ in reduced units, electrons coherently traversing 
the QPC do not carry enough energy to emit photons to the cavity and the cavity remains in the vacuum state at zero temperature. 
Despite this absence of photons, vacuum fluctuations in the cavity still affect and suppress electron tunneling with Frank-Condon factors~\cite{BergenfeldtSET2012}. 
This suppression is alternatively captured by the equilibrium $P(E)$ theory~\cite{ingold-nazarov-book1992} of the DCB \cite{yeyati-urbina-prl2001,golubev-zaikin-prl2001}. 

For $V> \hbar \omega_0/e$, the electrons scattered by the QPC emit and absorb photons to/from the cavity. We want to characterize the distribution of photons by studying the mean number of photon, the damping rate of the cavity and the second-order coherence $g^{(2)} (0)$ at vanishing time. For a weak electron-photon (e-p) coupling, the field radiated in an open transmission line exhibits~\cite{beenakker-schomerus-prl2001} a negative-binomial form for the photon distribution, similar to blackbody radiation, approaching a Poisson distribution with $g^{(2)} (0)=1$ when the number of bosonic modes is infinite. In our case of a single-mode cavity, we consistently find a thermal distribution with $g^{(2)} (0)=2$ as with other conductors~\cite{BergenfeldtSET2012,jin-schon-prb2015}, independently of the transmission coefficients of the QPC. Proceeding with the next-to-leading order at weak light-matter coupling, we obtain analytically a decrease in $g^{(2)} (0)$ controlled by the bias voltage $V$. This decrease is caused by a two-photon absorption process whereas two-photon emission is energetically forbidden for $V < 2 \hbar \omega_0/e$. The balance is reestablished at large bias where $g^{(2)} (0)=2$ is recovered.

Interestingly, our next-to-leading order calculation can also be understood as a backaction effect. In the presence of electron transport accompanied by photon emission, the cavity reaches a non-equilibrium stationary state characterized by a Bose-Einstein photon distribution. This cavity state generates an out-of-equilibrium DCB affecting transport which in turn modifies the photon distribution. Here we show that the conventional $P(E)$ theory extended to the stationary state~\cite{Souquet2014} recovers quantitatively the results of the straightforward perturbative calculation, revealing further the physics behind our analytical results. 

We also discuss the connections between different theoretical approaches to describe the coupled QPC-cavity system. We first consider a capacitive model initially devised in Ref.~\onlinecite{dmytruk-mora-simon-prb2015} and included here in the Keldysh path integral framework~\cite{kamenev2011,torre2013}. By inspecting the leading electron-photon order, we show its equivalence with a Keldysh path integral method introduced by Kindermann and Nazarov~\cite{kindermann2003,snyman-nazarov-prb2008} in which electronic variables have already been integrated out. We also apply a rate equation approach, obtained by neglecting off-diagonal elements in the quantum master equation, and show that it coincides with results from the path integral methods in the rotating-wave approximation (RWA). Denoting $\kappa$ the damping rate of the cavity, we focus below on the case $|eV - \hbar \omega_0| \gg \kappa$ where RWA is applicable, leaving aside the regime $eV \simeq \hbar \omega_0$ where the antibunching inherited from electrons plays a crucial role in the photon distribution~\cite{hassler-otten-prb2015}.

The article is organized as follows. Section~\ref{qpc-tl} discusses different architectures coupling the QPC to the TLC. In Sec.~\ref{sec1} we introduce the capacitive model and derive an effective action for the photons at weak e-p coupling. We compute the number of photon, the cavity damping rate, the frequency pull and the second-order coherence $g^{(2)}(\tau)$. The alternative action form in which the electron degrees of freedom are exactly integrated is introduced in Sec.~\ref{sec2} and the previously derived effective action at weak e-p coupling is recovered. Further assuming small transmissions $T_{n} \ll 1$, we determine the first non-quadratic corrections to the effective action and provide results for the average number of photons, the cavity damping rate and $g^{(2)}(0)$. These results are exactly recovered in Sec.~\ref{sec3} using rate equations and in Sec.~\ref{sec4} by implementing a non-equilibrium $P(E)$ theory to describe the cavity DCB. Sec.~\ref{concl} concludes.

\section{QPC-TLC coupling \label{qpc-tl}}
%%%%%%%%%%%%%%%%%%%%%%%%%%%%%%%%%%% Figure 1 %%%%%%%%%%%%%%%%%%%%%%%%%%%%%%%%%%%%%%%%%%%%
\begin{figure}[!h]
\begin{center}
\includegraphics[width=0.4\textwidth]{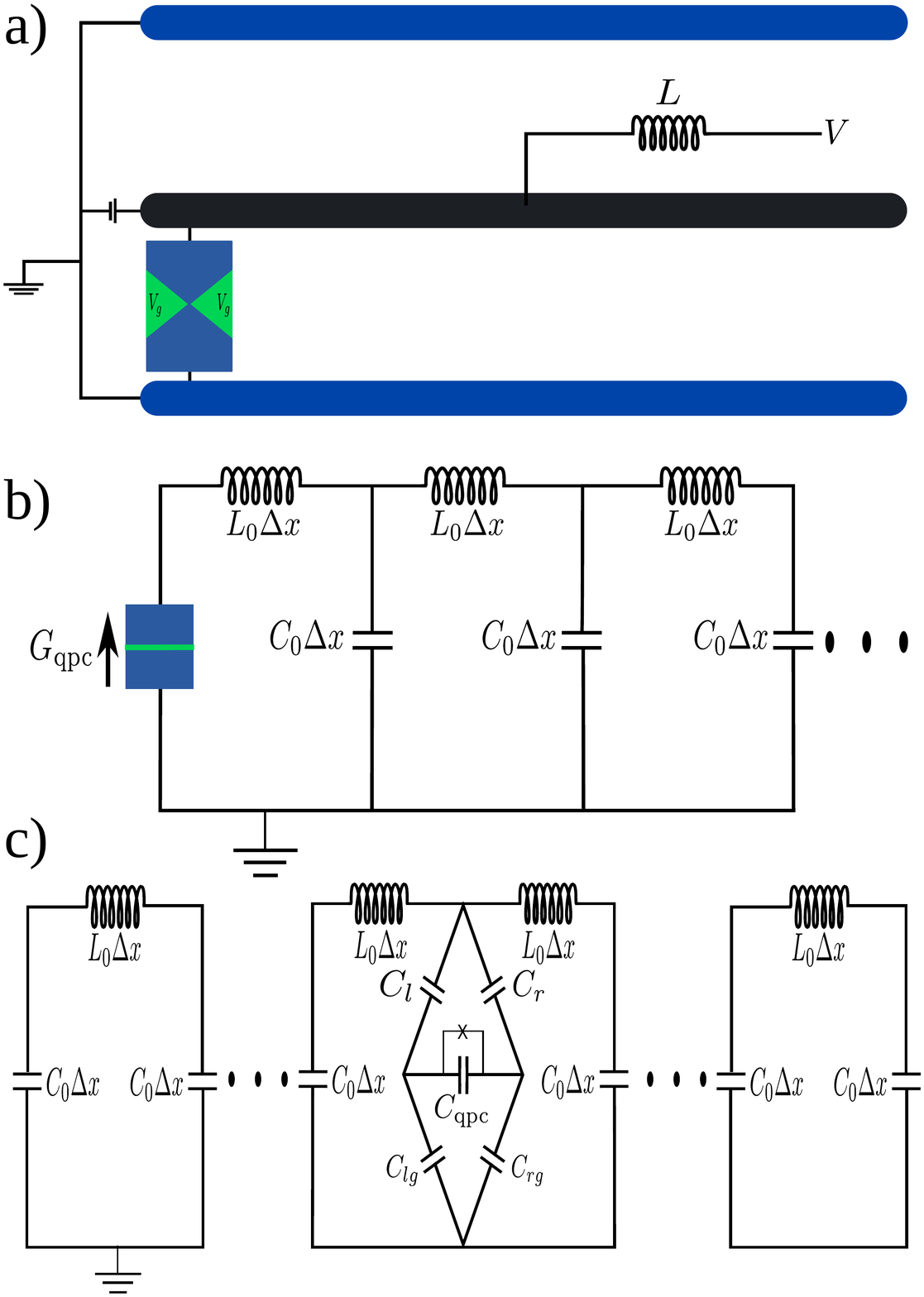}
\caption{(a) Schematic representation of the TLC coupled to the QPC. The QPC is defined by the gate voltage $V_{g}$ that controls the transmission probability 
$T_{n}$ of each channels $n$. A voltage $V$ applied to the TLC voltage node controls the electronic transport in the QPC. (b) and (c) Circuit representation 
of a galvanic and capacitive coupling between the TLC and the QPC, respectively. \label{fig1}}
\end{center}
\end{figure}
%%%%%%%%%%%%%%%%%%%%%%%%%%%%%%%%%%% Figure 1 %%%%%%%%%%%%%%%%%%%%%%%%%%%%%%%%%%%%%%%%%%%%

Here we discuss how the QPC can be coupled to a TLC. Fig.~\ref{fig1}(a) depicts the TLC-QPC hybrid system. The TLC is characterized by its length 
$d$ and frequency $\omega_{0}=1/\sqrt{C_{c}L_{c}}$, where $C_{c} = C_{0}d$ and $L_{c}=L_{0}d$ are the TLC capacitance and inductance, respectively. 
The QPC can be coupled to the TLC either via a galvanic \cite{altimiras2014,Parlavecchio2015} or capacitive coupling \cite{Petersson2012,frey-wallraff-prl2012}. 
In the galvanic coupling scheme, Fig.~\ref{fig1}(b), the TLC is terminated by the QPC and the dimensionless coupling constant is 
\begin{equation}
\lambda_g = \sqrt{\frac{\pi Z_\text{c}}{R_\text{K}}}
\end{equation}
where $Z_\text{c}=\sqrt{L_c/C_c}$ is the cavity impedance and $R_{\text{K}}=h/e^2$ is the quantum resistance. A coupling strength of $\lambda_{g}^{2} \approx 0.3$ 
has been reached~\cite{altimiras2014} and even larger values are expected in a near future~\cite{samkharadze2015}.

In the capacitive coupling scheme, Fig.~\ref{fig1}(c), the QPC is inside the cavity and both left ($l$) and right ($r$) electronic 
reservoir are coupled to the central and ground lines via capacitances $C_{\alpha}$ and $C_{\alpha g}$, where $\alpha=l,r$. The reservoirs are also coupled 
between themselves via capacitance $C_\text{qpc}$. In this case, the dimensionless coupling constant between the lead $\alpha$ and the cavity field is
\begin{align}
\lambda_{\alpha} = \frac{C_\alpha C_{\alpha^{\prime} s} + C_\text{qpc}C_s}{C_{ls}C_{rs}+C_\text{qpc}C_s}\frac{ef(x_0)}{\sqrt{2\hbar\omega_{0}C_{c}}}
\end{align}
where $\alpha^{\prime}=l,r$ with $\alpha\neq \alpha^{\prime}$, $e$ is the electron charge, $f(x_0)$ is the wavefunction of the cavity field at the coupling 
position $x_0$, $C_{\alpha s}=C_\alpha+C_{\alpha g}$ and $C_s=C_{ls}+C_{rs}$. The coupling constant was obtained via circuit theory \cite{koch_girvin_pra2010,BergenfeldtSET2012}. 
A detailed description of the circuit theory applied to a quantum dot coupled to the TLC is given in Ref.~[\onlinecite{BergenfeldtSET2012}]. 

In the next section, we show that the photonic properties are characterized by the difference between the e-p coupling of each lead with the TLC, 
i.e., the relevant quantity defining the coupling is 
\begin{equation} \label{eq-lamb1}
\lambda_c = \lambda_{l}- \lambda_{r}= \frac{C_l C_{rg}-C_r C_{lg}}{C_{ls}C_{rs}+C_\text{qpc}C_s}\frac{ef(x_0)}{\sqrt{2\hbar\omega_{0}C_{c}}}.
\end{equation}
This equation reveals that if $C_{l(g)}=C_{r(g)}$ the QPC is completely decoupled from the TLC. It also shows that the maximum coupling occurs when one 
reservoir is coupled to the central line and the other one to ground line. Thus, in this geometry, placing the QPC where the cavity field is maximum 
and considering the typical values for the capacitances $C_l \approx C_{rg} \approx 10$ fF and $C_\text{qpc} \approx 0.1$ fF we have 
that $\lambda_{c} \approx 0.98 \lambda_{g}$. In this geometry the capacitive coupling strength is of the same order as the galvanic one.

It is interesting to note that $\lambda_{g(c)}^{2}\hbar\omega_{0}=E_{c_{g(c)}}$, where $E_{c_{g(c)}}$ is the cavity charging energy. The different types 
of coupling the QPC to the TLC produce different charging energies. For the galvanic coupling $E_{c_{g}}=e^{2}/2C_{c}$, while in the capacitive one 
$E_{c_{c}}=e^{2}/2C_\text{eff}$, where $C_\text{eff}=C_{c}[C_{ls}C_{rs}+C_\text{qpc}C_s]^{2}/[(C_l C_{rg}-C_r C_{lg})f(x_0)]^2$. 

From now on we use $\lambda$ to refer to the dimensionless e-p coupling and $E_{c}$ to the charging energy independently of the way the cavity is coupled 
to the QPC.

\section{Model and Keldysh path-integral \label{sec1}}

\subsection{Formalism}

We consider a single-mode cavity field coupled to both sides of the constriction with different coupling constants $g_{l(r)}=\hbar\omega_{0} \lambda_{l(r)}$. 
The hybrid system Hamiltonian is
\begin{equation} \label{Hamil}
\hat{H} = \hbar\omega_{0}\hat{a}^{\dagger}\hat{a} + \hat{H}_{\rm qpc} + (\hat{a}^{\dagger}+\hat{a})\hat{\eta}.
%(g_{\text{L}}\hat{n}_{\text{L}}+g_{\text{R}}\hat{n}_{\text{R}}),
\end{equation}
The first term is the single-mode cavity field Hamiltonian, $\omega_{0}$ is the photon frequency and $\hat{a}$ ($\hat{a}^{\dagger}$) the photon annihilation 
(creation) operator. The second term is the Hamiltonian describing the QPC electronic degrees of freedom, it contains the kinetic 
energy of the electronic leads, the scattering term between them and electronic interactions. The last term describes the TLC-QPC interaction, where $\hat{\eta}=g_l\hat{n}_l+g_r\hat{n}_r$ and $\hat{n}_{\alpha}$ is the operator giving the total number of electrons in the lead $\alpha$. 
A rigorously description of the e-p coupling via circuit theory gives rise to a capacitive electron-electron interaction \cite{BergenfeldtSET2012}. 
We neglected this term in our description.

We use the Keldysh path-integral formalism to investigate the cavity field properties \cite{torre2013,kamenev2011}. The partition function 
describing the cavity-QPC hybrid device is
\begin{align}
\mathcal{Z} &= \int \mathcal{D}[a,a^{*},c,c^{*}]e^{i (\mathcal{S}_{\text{cav}}+ \mathcal{S}_{\text{e-p}}+ \mathcal{S}_{\text{qpc}})/\hbar},
\end{align}  
where $a$ ($c$) is the complex photonic (fermionic) field. $\mathcal{S}_{\text{cav}}$, $\mathcal{S}_{\text{e-p}}$ and $\mathcal{S}_{\text{qpc}}$ 
are, respectively, the cavity field, cavity-QPC interaction and QPC actions. In the frequency-domain the cavity action, in terms of the classical 
and quantum components of the field $a$, is written as
\begin{equation} \label{Scav}
\mathcal{S}_{\text{cav}} = \int_{\omega} (a_{c}^{*},a_{q}^{*})_{\omega} \begin{pmatrix}
0 & [G_{0}^{\text{R}}]^{-1}(\omega) \\ [G_{0}^{\text{A}}]^{-1}(\omega) & [G_{0}^{\text{K}}]^{-1}(\omega) \end{pmatrix} \begin{pmatrix}
a_{c} \\ a_{q} \end{pmatrix}_{\omega},
\end{equation}
where we use the notation $\int_{\omega} = \int d\omega /2\pi$. $[G_{0}^{\text{R(A)}}]^{-1}(\omega) = \hbar\omega - \hbar\omega_{0} \pm i\hbar 0$ 
is the bare retarded (advanced) Green's function (GF), and the bare Keldysh GF $[G_{0}^{\text{K}}]^{-1}(\omega)$ is only a regularization, which 
is neglected in the presence of the self-energies.

The cavity-QPC action is
\begin{equation} \label{epaction}
\mathcal{S}_{\text{e-p}}=-\int \frac{dt}{\sqrt{2}} \{[a_{c}(t)+a_{c}^{*}(t)]\eta_{q}(t)+[a_{q}(t)+a_{q}^{*}(t)]\eta_{c}(t)\}
\end{equation}
with $\eta_{c(q)}(t) = \eta_{+}(t) \pm \eta_{-}(t)$, $\eta_{\pm}(t)= \sum_{\alpha = l,r} g_{\alpha}n_{\alpha}^{\pm}(t)$, and $n_{\alpha}^{\pm}(t)$ 
is the density field in the forward ($+$) and backward ($-$) branches of the Keldysh time-ordered contour. Since we are interested in the cavity 
field properties the specific formula of the QPC action is not necessary and, hence, we keep it as general as possible. The QPC is only characterized 
by its transmission probability $T_{n}$.

Integrating over the fermionic degrees of freedom obtains an effective action ($\mathcal{S}_{\text{eff}}$) describing the photons. Assuming a small 
cavity-QPC coupling constant we derive $\mathcal{S}_{\text{eff}}$ using the cumulant expansion~\cite{mendes-mora-njp2015}. To second-order in the e-p 
coupling we obtain after averaging over electrons
\begin{equation}
\langle e^{i \mathcal{S}_{\text{e-p}}/\hbar}\rangle_{e} \simeq  e^{i (\langle \mathcal{S}_{\text{e-p}}\rangle_{e} 
+i \langle \delta \mathcal{S}_{\text{e-p}}^{2}\rangle_{e}/ 2\hbar)/\hbar},
\end{equation}
with $\langle \ldots \rangle_{\text{e}} = \int \mathcal{D}[c,c^{*}](\ldots) e^{i \mathcal{S}_{\text{qpc}}/\hbar}$, and $ \delta \mathcal{S}_{\text{e-p} } 
= \mathcal{S}_{\text{e-p} } - \langle \mathcal{S}_{\text{e-p} } \rangle_{\text{e}}$. Thus, the photon effective action takes the form
\begin{equation}
\mathcal{S}_{\text{eff}} = \mathcal{S}_{\text{cav}} + \langle \mathcal{S}_{\text{e-p}}\rangle_{e} + \frac{i}{2\hbar}\langle \delta \mathcal{S}_{\text{e-p}}^{2}\rangle_{e}.
\end{equation}
The first term describes the uncoupled TLC, the second and third terms are the linear and quadratic contribution resulting from the e-p coupling. 
The linear term is
\begin{align} \label{s-linear}
\langle \mathcal{S}_\text{e-p} \rangle_{\text{e}} &= -\sqrt{2}\int_{\omega} [a_{q}^{*}(\omega)+a_{q}(-\omega)][g_{r} \langle \hat{n}_{r}(\omega) \rangle + g_{l}\langle \hat{n}_{l}(\omega)\rangle]
\end{align}
with $\langle \hat{n}_{\alpha}(\omega) \rangle = \langle n_{\alpha}^{+}(\omega) \rangle_{e}= \langle n_{\alpha}^{-}(\omega) \rangle_{e}$. As expected it depends 
only on the quantum fields, since a dependence on classical fields would violate the causality structure~\cite{kamenev2011}.

The quadratic action
\begin{align} \label{Sep2}
\langle \delta \mathcal{S}_{\text{e-p} }^{2} \rangle_{\text{e}} & = - \frac{2\hbar}{i}\int_{\omega} (a_{c}^{*},a_{q}^{*})_{\omega}
\begin{pmatrix} 0 & \Sigma^{\text{A}}(\omega) \\ \Sigma^{\text{R}}(\omega) & \Sigma^{\text{K}}(\omega) \end{pmatrix}  
\begin{pmatrix} a_{c} \\ a_{q} \end{pmatrix}_{\omega} \nonumber \\
&- \mathcal{S}_{\text{a}}
\end{align}
introduces the advanced $\Sigma^\text{A}(\omega)$, retarded $\Sigma^\text{R}(\omega)$ and Keldysh $\Sigma^\text{K}(\omega)$ self-energies, while
\begin{equation*} \label{Sa}
\mathcal{S}_\text{a} = \frac{2\hbar}{i}\int_{\omega} [a_{c}(-\omega)a_{q}(\omega)\Sigma^{\text{R}}(\omega) + a_{q}(-\omega)a_{q}(\omega)\frac{\Sigma^{\text{K}}(\omega)}{2}+ \text{c.c}],
\end{equation*}
is the anomalous action. This term is responsible for producing quadrature squeezing when the quantum conductor is ac-biased~\cite{mendes-mora-njp2015}. 
In the time-domain the retarded and Keldysh self-energies are expressed as
\begin{subequations} \label{retard}
\begin{align} 
\Sigma^{\text{R}}(t_{2}-t_{1}) &= -\frac{i}{\hbar} \Theta(t_{2}-t_{1})\langle [\delta \hat{\eta}(t_{2}), \delta \hat{\eta}(t_{1})]\rangle \\
\Sigma^{\text{K}}(t_{1}-t_{2}) &= -\frac{i}{\hbar}\langle \{\delta \hat{\eta}(t_{1}), \delta \hat{\eta}(t_{2})\}\rangle. 
\end{align}
\end{subequations}
with the notation $\delta \hat{A} (t) \equiv \hat{A} (t) - \langle \hat{A} \rangle$, thus in terms of the density-density correlators
\begin{equation} \label{dens_corr}
\langle \delta\hat{\eta}(t_{1})\delta\hat{\eta}(t_{2}) \rangle = \sum_{\alpha,\beta =l,r} g_{\alpha} g_{\beta} \langle \delta \hat{n}_{\alpha}(t_{1})\delta \hat{n}_{\beta}(t_{2}) \rangle.
\end{equation}
Following Ref.~\onlinecite{dmytruk-mora-simon-prb2015}, it is convenient to rewrite these correlators using current operators
\begin{equation} \label{dens-curr}
\hat{n}_{\alpha}(t) = \frac{i}{e}\int_{\omega} \frac{\hat{I}_{\alpha}(\omega)}{\omega} e^{-i\omega t},
\end{equation}
where $\hat{I}_{\alpha}$ is the charge current towards the lead $\alpha = l/r$. For a time-independent bias applied to the QPC, we introduce 
the noise power spectrum function $\langle \delta \hat{I}_{\alpha}(\omega_{1}) \delta \hat{I}_{\beta}(\omega_{2}) \rangle = 2\pi S_{\alpha \beta}(\omega_{1}) \delta(\omega_{1}+\omega_{2})$ in its non-symmetrized form. Using the above relations, the self-energies of Eqs.~\eqref{retard} 
decompose in frequency space as
\begin{align*}
\Sigma^{\text{R}}(\omega) =\Delta(\omega)-i\Gamma(\omega)/2, \qquad \Sigma^{\text{A}}(\omega)=\Sigma^{\text{R} *}(\omega)
\end{align*}
with the real part
\begin{align} \label{real_part}
\Delta(\omega) &= -\sum_{\alpha,\beta} \frac{g_{\alpha}g_{\beta}}{e^2\hbar} \mathcal{P}\int_{\omega_{1}} \frac{S_{\alpha \beta}(\omega_{1})-
S_{\alpha \beta}(-\omega_{1})}{\omega_{1}^{2}(\omega_{1}+\omega)} 
\end{align}
and the imaginary part
\begin{align} \label{imag_part}
\Gamma(\omega)&=\sum_{\alpha,\beta} \frac{g_{\alpha}g_{\beta}}{e^2\hbar} \frac{S_{\alpha \beta}(\omega)-S_{\alpha \beta}(-\omega)}{\omega^{2}},
\end{align}
while the Keldysh component is 
\begin{align*}
\Sigma^\text{K}(\omega) = -i\sum_{\alpha,\beta} \frac{g_{\alpha}g_{\beta}}{e^2\hbar} \frac{S_{\alpha \beta}(\omega)+S_{\alpha \beta}(-\omega)}{\omega^{2}}.
\end{align*}
The photon self-energies are completely characterized by the non-symmetrized auto- ($\alpha=\beta$) and cross-correlation 
($\alpha\neq \beta$) noise spectra. While the retarded self-energy is given by the difference between absorption noise ($\omega>0$) 
and emission noise ($\omega<0$), the Keldysh component is proportional to the sum of them, {\it i.e}, the symmetrized noise.

The advantage of introducing charge currents instead of densities in Eqs.~\eqref{retard} is that the resulting noise correlators can be computed by various techniques, in particular the conventional Landauer-B\"uttiker scattering formalism~\cite{blanter2000} in the case of a QPC. Moreover noise correlators are measurable quantities which can be accessed experimentally.

As the self-energies are proportional to the square of the e-p coupling constant the pole of the GFs are weakly modified by them. Therefore, we 
approximate $\Sigma^{i}(\omega) \approx \Sigma^{i}(\omega_{0})$, with $i=\text{R, A, and K}$. This approximation is equivalent to the rotating-wave 
approximation (RWA) performed in the time-domain consisting in averaging to zero all the fast oscillating terms.

The anomalous action, $\mathcal{S}_\text{a}$, is neglected within the RWA, since its contribution oscillates with frequency $2\omega_{0}$. 
The final expression for the effective action is obtained by shifting the classical field to absorb the linear action 
$\langle \mathcal{S}_{\text{e-p}} \rangle_{\text{e}}$, namely
\begin{equation}
a_{c}(\omega) \rightarrow a_{c}(\omega) + \sqrt{2}\sum_{\alpha}g_{\alpha} \langle \hat{n}_{\alpha}(\omega)\rangle G_{\text{R}}(\omega),
\end{equation}
thereby producing a correction to the photons correlators. For the number of photons the correction is proportional to $(\lambda \sum_n T_n/R_\text{K})^2$, 
which is higher-order in the e-p coupling and can be neglected, see below. The third and fourth orders in the e-p 
coupling discussed in Sec.~\ref{sec2} are also not altered by this shift in the classical field, as discussed in Appendix~\ref{appA}.

\subsection{Results from scattering theory \label{stheory}}

The photon effective action, given by Eq.~\eqref{Scav} and the first term of Eq.~\eqref{Sep2}, is 
\begin{align} \label{Seff}
\mathcal{S}_{\text{eff}} & = \int_{\omega} (a_{c}^{*},a_{q}^{*})_{\omega} \begin{pmatrix} 0 & G_{\text{A}}^{-1}(\omega) \\ G_{\text{R}}^{-1}(\omega) 
& -\Sigma^{\text{K}}(\omega_{0}) \end{pmatrix}  \begin{pmatrix} a_{c} \\ a_{q} \end{pmatrix}_{\omega}
\end{align}
where $G_{\text{R(A)}}^{-1}(\omega) = \hbar\omega - \hbar\omega_{0} - \Sigma^{\text{R(A)}}(\omega_{0})$ is the inverse of the retarded (advanced) 
GF. The Keldysh GF is $G^{K}(\omega) = G_{\text{R}}(\omega)\Sigma^{\text{K}}(\omega_{0})G_{\text{A}}(\omega)$. So far the photon effective action 
has been derived assuming only weak electron-photon coupling but arbitrary transmission or electron-electron interactions.

To further characterize the photon effective action, we assume a non-interacting QPC, i.e., in the absence of both electron-electron interaction 
and e-p coupling, which implies in energy-independent transmission probabilities \cite{blanter2000}. We compute the nonsymmetrized noise via the scattering 
formalism~\cite{aguado-kouwenhoven-prl2000,blanter2000}. Charge conservation which imposes
\begin{equation}
S_{rr}(\omega)= S_{ll}(\omega)=-S_{rl}(\omega)=-S_{lr}(\omega). 
\end{equation} 
with
\begin{align} \label{qpc-nsnoise}
S_{rr}(\omega) &= \frac{1}{R_{\text{K}}}\left[2\hbar\omega \Theta(\hbar\omega) \sum_{n} T_{n}^{2}+\sum_{n} T_{n}R_{n}\bar{S}(\omega)\right],
\end{align}
where we introduce the notation
\begin{equation}\label{eq-noise}
\bar{S}(\omega)=(\hbar\omega+eV)\Theta(\hbar\omega+eV)+(\hbar\omega-eV)\Theta(\hbar\omega-eV),
\end{equation}
$\Theta(\omega)$ is the Heaviside step function and $R_n=1-T_n$. 

We are now in a position to compute the the real and imaginary parts of the retarded self-energy. The real part 
\begin{equation}
\Delta(\omega) = -\frac{(g_r-g_l)^{2}}{\hbar\pi}\sum_{n}T_{n} \mathcal{P}\int_{\omega_{1}}\frac{1}{\omega_{1}(\omega_{1}+\omega)}=0,
\end{equation}
is proportional to the difference between the e-p couplings and the transmission probability. However, the integral is zero for any value of 
$\omega$, meaning that the cavity frequency is not shifted by the e-p coupling. This result extends the absence of a cavity frequency pull 
obtained for a tunnel junction~\cite{mendes-mora-njp2015,dmytruk-mora-simon-prb2015} to leading order in the e-p coupling. Higher orders in 
the coupling or an energy dependence of the transmission provide the non-linearities needed for a shift in the cavity resonant frequency.

The imaginary part of the self-energy is related to the exchange of photons between the cavity and the QPC. It determines the cavity damping 
rate $\kappa$, i.e, the photon losses to the electronic environment. To leading order, $\kappa \simeq \kappa_0 = \Gamma(\omega_{0})/\hbar$, 
with 
\begin{equation} \label{kapp}
\kappa_0 =\frac{\lambda^{2}}{e^2}[S_{rr}(\omega_{0})-S_{rr}(-\omega_{0})] = \frac{\omega_{0}\lambda^{2}}{\pi}\sum_{n} T_{n},
\end{equation}
where $\lambda=(g_r -g_l)/\hbar\omega_{0}$. The cavity damping rate has a simple dependence on the transmission probability $T_{n}$, 
and it increases with the number of electron channels. Considering a single channel QPC and $T_{1}=1$, one can access the dimensionless 
e-p constant by measuring the cavity peak broadening \cite{Petersson2012,frey-wallraff-prl2012,toida2013,Deng-leHur-guo-arxiv2015}.

Finally, the Keldysh self-energy 
\begin{align*}
\Sigma^{\text{K}}(\omega_{0}) = -i\frac{\lambda^{2}R_{\text{K}}}{2\pi}[S_{rr}(\omega_{0})+S_{rr}(-\omega_{0})],
\end{align*}
also depends on the QPC transmissions. We note that the different self-energies are non-zero only if $g_r$ differs from $g_l$ corresponding 
to an inhomogeneous coupling between the leads and the cavity field.

\subsection{Cavity field properties}

We now present results for the cavity photons. The number of photons is related to the Keldysh GF via the formula 
$2 \langle n \rangle +1=iG^\text{K}(t=0)$. From the Gaussian action~\eqref{Seff}, we easily find 
\begin{align} \label{nphotons}
\langle n \rangle &= \frac{S_{rr}(-\omega_{0})}{S_{rr}(\omega_{0})-S_{rr}(-\omega_{0})}= F\frac{(eV-\hbar\omega_{0})}{2\hbar\omega_{0}}\Theta(eV-\hbar\omega_{0}),
\end{align}
where $F=\sum_{n}T_{n}(1-T_{n})/\sum_{n}T_{n}$ is the QPC Fano factor. The number of photons is determined by the emission noise~\cite{beenakker-schomerus-prl2004,LoosenLesovikJETP1997} over 
the rate at which the cavity loses photons to the QPC. At zero temperature and $V \leq \hbar \omega_{0}/e$ the number of photons is zero. 
In this case, electrons traversing the QPC must have an energy in a window of size $e V$ and are therefore not able to emit a photon 
with energy $\hbar \omega_0$ to the cavity. Remarkably, the number of photons decreases as the transmission probabilities increase and 
even vanish in the limit of perfectly transmitting channels. The physical reason is that photon emission, similarly to quantum noise, 
is related to charge discreteness in electron transport. At perfect transmission, there is a continuous flow of charges with no noise 
and no photon emission. In this case however, the damping rate of the cavity $\kappa_0$ remains finite as the bath of electrons can 
still absorb photons from the cavity. For weak transmissions $T_n \ll 1$, the result~\eqref{nphotons} coincides 
with previous studies for a tunnel junction~\cite{jin-schon-prb2015,mendes-mora-njp2015} and metallic QDs\cite{BergenfeldtSET2012}.

Next we compute the photon second-order coherence $g^{(2)}(\tau)=\langle \hat{a}^{\dagger}(0)\hat{a}^{\dagger}(\tau)\hat{a}(\tau)\hat{a}(0) \rangle/\langle n \rangle^{2}$. 
At zero time, $g^{(2)}(0)=\langle n (n-1) \rangle/\langle n \rangle^{2}$ indicates whether the cavity field presents super-Poissonian 
($g^{(2)}(0) >1$) or sub-Poissonian ($g^{(2)}(0) <1$) photon statistics. Also, its time-evolution is a direct measurement of photon bunching 
($g^{(2)}(\tau)< g^{(2)}(0)$) or antibunching ($g^{(2)}(\tau)> g^{(2)}(0)$). The quadratic action~\eqref{Seff} implies a Gaussian field 
distribution, similar to a thermal state, for which the calculation of correlation functions is straightforward by using Wick's theorem. 
$g^{(2)}(\tau)$ is thus related to the first-order coherence $g_1 (\tau) = \langle \hat{a}^{\dagger}(\tau) \hat{a}(0) \rangle/\langle n \rangle$ 
whose modulus follows a simple exponential decay with the cavity damping rate $\kappa_0$. The result
\begin{equation}
g^{(2)}(\tau)= 1 + | g_1 (\tau) |^2 = 1+e^{-\kappa_0 \tau},
\end{equation} 
exhibits photon bunching and super-Poissonian statistics $g^{(2)}(0) = 2$. These results are in fact a direct consequence of the Gaussian 
field distribution~\cite{BergenfeldtSET2012,jin-schon-prb2015}. They are valid at weak coupling $\lambda \ll 1$ where non-quadratic terms 
in the effective action can be discarded. The case of a dc-bias Josephson junction shows that increasing e-p interaction may lead to strongly 
antibunched photons~\cite{dambach2015}. The next section is devoted to the study of non-quadratic terms in the action 
to investigate in particular how they modify $g^{(2)} (0)$ as the coupling $\lambda$ increases.

\section{Beyond the thermal distribution \label{sec2}}

\subsection{Alternative gauge}

We showed above that the cavity field follows a thermal distribution at weak electron-photon coupling. Deviations to this statistical 
description emerge by expanding the action~\eqref{epaction} beyond second order. Here, instead of expanding further Eq.~\eqref{epaction}, \cite{}
we use an alternative gauge for which the action integrated over electronic variables has been derived exactly~\cite{kindermann2003,snyman-nazarov-prb2008}. 
This model was recently used to study a quantum tunneling detector coupled to a coherent conductor \cite{tobiska-nazarov-2006} and the light 
emitted by the tunneling of electrons from a STM tip to a metallic surface \cite{xu-belzig-prl2014}.

The corresponding action is $\mathcal{S}_{\text{cav}}+ \mathcal{S}_{\text{e-p}}$ where the unperturbed cavity action $\mathcal{S}_{\text{cav}}$ 
is still given by Eq.~\eqref{Scav}. The average over electrons produces the non-linear photon action
\begin{equation} \label{actqpc2}
\mathcal{S}_{\text{e-p}} = -\frac{i\hbar}{2}\sum_{n} \Tr \ln \left[1 + \frac{T_{n}}{4}\left(\{ \check{G}_l,\check{G}_r\}-2\right)\right],
\end{equation}
where the trace is over the Keldysh indices ($\pm$) and time. $\check{G}_{\alpha}$ is the Keldysh GF of the electrons in the reservoir $\alpha$, 
and they are defined in terms of the equilibrium GF
\begin{equation} \label{eqKGF}
\check{G}_{\text{eq}}^{\alpha}(\epsilon) = \begin{pmatrix}
1-2f(\epsilon_\alpha) & 2f(\epsilon_\alpha) \\
2-2f(\epsilon_\alpha) & 2f(\epsilon_\alpha)-1
\end{pmatrix},
\end{equation}
where $f(\epsilon)$ is the Fermi function. $\check{G}_{r}(t,t^{\prime})= \check{G}_{\text{eq}}^{r}(t-t^{\prime})$ and 
$\check{G}_{l}(t,t^{\prime}) = \check{U}^{\dagger}(t)\check{G}_{\text{eq}}^{l}(t-t^{\prime})\check{U}(t^{\prime})$ with
\begin{equation} \label{gauge-eq}
\check{U}(t) = \begin{pmatrix} e^{\lambda \varphi_{+}(t)} && 0 \\ 0 && e^{\lambda \varphi_{-}(t)}\end{pmatrix}
\end{equation} 
where $\varphi_{\pm}(t)=a_{\pm}^{*}(t)-a_{\pm}(t)$ and $a_{\pm}$ is the complex photon field defined in the $\pm$ Keldysh contour. The matrix 
$\check{U}(t)$ describes the photons. For details on the derivation of $\mathcal{S}_\text{e-p}$ see Ref. [\onlinecite{snyman-nazarov-prb2008}].

The link between this description and the model from Sec.~\ref{sec1} is best seen in the tunneling limit where scattering in the QPC is accounted 
for by the tunnel Hamiltonian $\hat{H}_{T} = \hat{\mathcal{T}}+\hat{\mathcal{T}}^{\dagger}$, where $\hat{\mathcal{T}}$ describes the tunneling of 
one electron from the left to the right reservoir and $\hat{\mathcal{T}}^{\dagger}$ the reversed process. Applying the gauge transformation via 
the unitary operator $\hat{U}_0 = \exp[\hat{\eta} (\hat{a}-\hat{a}^\dagger)/\hbar\omega_{0}]$ cancels the electron-photon coupling term 
$(\hat{a}^{\dagger}+\hat{a})\hat{\eta}$ in Eq.~\eqref{Hamil} while dressing the tunneling part $\bar{H}_{T}= \hat{U}_0^{\dagger}\hat{H}_{T} \hat{U}_0 = \hat{\mathcal{T}}e^{-\lambda(\hat{a}^{\dagger}-\hat{a})}+\hat{\mathcal{T}}^{\dagger}e^{\lambda(\hat{a}^{\dagger}-\hat{a})}$ with 
$\lambda = (g_{r}-g_{l})/\hbar\omega_{0}$. This new form of the tunnel Hamiltonian implies that each tunneling event is accompanied by 
the coherence excitation of the cavity with the displacement operators $e^{\pm \lambda(\hat{a}^{\dagger}-\hat{a})}$. The same prescription 
was used in deriving~\cite{snyman-nazarov-prb2008} Eq.~\eqref{actqpc2}. For completeness, we also show in Appendix~\ref{appA} that the 
expansion of Eq.~\eqref{actqpc2} to second order in $\lambda$ agrees with the results of the previous section with the same identification 
$\lambda = (g_{r}-g_{l})/\hbar\omega_{0}$.

\subsection{Non-quadratic effects \label{nq-action}}

We assume for simplicity weak transmission probabilities $T_{n}\ll 1$ and expand the action~\eqref{actqpc2} as
\begin{equation} \label{actqpc2ex}x
\mathcal{S}_{\text{e-p}} \simeq -\frac{i\hbar g_c}{8}\sum_{n} \Tr \left[ \left(\{ \check{G}_l,\check{G}_r\}-2\right)\right],
\end{equation}
where we introduce the dimensionless conductance $g_c  =\sum_{n}T_{n}$. In this limit, the QPC description is equivalent to a tunnel junction.

Equation~\eqref{actqpc2ex} is further expanded to fourth order in $\lambda$. The effective action is
\begin{equation} \label{seff2}
\mathcal{S}_{\text{eff}} = \mathcal{S}_{\text{q}} + \mathcal{S}_{\text{nq}}^{(3)} + \mathcal{S}_{\text{nq}}^{(4)}.
\end{equation}
 The second order $\mathcal{S}_{\text{q}}$ (from now on, the subscript q stands for the 
quadratic approximation) has been derived in the previous section, see Eq.~\eqref{Seff}, and is rederived in Appendix~\ref{appA} 
for the present model. Assuming weak transmission implies a Fano factor $F=1$ in Eq.~\eqref{nphotons} for the average number of 
photons $\langle n \rangle_\text{q}$, or
\begin{equation}\label{nq}
\langle n \rangle_\text{q} = \frac{\lambda^2 S_{rr} (-\omega_0)}{e^2 \kappa_0} = \frac{(eV-\hbar\omega_0)}{2\hbar\omega_{0}}\Theta(eV-\hbar\omega_{0}).
\end{equation}
In this limit, the non-symmetrized noise power can be expressed as $S_{rr} (\omega) = (g_c/R_\text{K}) \bar{S}(\omega)$
where $\bar{S}(\omega)$ is given in Eq.~\eqref{eq-noise}.

The next terms in Eq.~\eqref{seff2} are derived in Appendix~\ref{appA}. The third-order expansion in the e-p coupling is
\begin{align} \label{eq-seff3}
&\mathcal{S}_\text{nq}^{(3)} =C_{0}\int dt_{1}dt_{2}d\epsilon_{l}d\epsilon_{r}
\sin[eV(t_{1}-t_{2})/\hbar] e^{-i\omega_{lr}(t_{1}-t_{2})} \nonumber \\
&\times f(\epsilon_{r})\bar{f}(\epsilon_{l})(2[\varphi_{+}(t_{2})-\varphi_{-}(t_{1})]^{3}-\mathsmaller{\sum}\limits_{\sigma}
[\varphi_{\sigma}(t_{2})-\varphi_{\sigma}(t_{1})]^{3})
\end{align}
where $C_0=-\lambda^{3}g_{c}/24\hbar\pi^2$, $\sigma=\pm$, $\bar{f}(\epsilon)=1-f(\epsilon)$ and $\omega_{lr}=(\epsilon_l-\epsilon_r)/\hbar$. 
Analogously, the fourth order term takes the form
\begin{align} \label{eq-seff4}
&\mathcal{S}_\text{nq}^{(4)} =C\int dt_{1}dt_{2}d\epsilon_{l}d\epsilon_{r}\cos[eV(t_{1}-t_{2})/\hbar] e^{-i\omega_{lr}(t_{1}-t_{2})} \nonumber \\
&\times f(\epsilon_{r})\bar{f}(\epsilon_{l})(2[\varphi_{+}(t_{2})-\varphi_{-}(t_{1})]^{4}-\mathsmaller{\sum}\limits_{\sigma}
[\varphi_{\sigma}(t_{2})-\varphi_{\sigma}(t_{1})]^{4}),
\end{align}
where $C =-i\lambda^{4}g_{c}/96\hbar\pi^{2}$. As the non-quadratic terms are small for $\lambda \ll 1$, we compute their contributions 
to the number of photons, retarded self-energy and $g^{(2)}(0)$ perturbatively. $\mathcal{S}_{\text{nq}}^{(3)}$ 
being odd in the number of bosons, it must be expanded at least to second order to contribute. The corresponding term is of order six 
in $\lambda$ and is negligible compared to $\mathcal{S}_{\text{nq}}^{(4)}$. It is discarded in what follows. 

The perturbation scheme employed here consists in expanding to first-order $e^{i\mathcal{S}_{\text{nq}}^{(4)}/\hbar} \approx 1 + i\mathcal{S}_{\text{nq}}^{(4)}/\hbar$. 
Any photon expectation value is obtained by computing
\begin{equation} \label{eqaverage}
\langle \ldots \rangle = \langle \ldots \rangle_\text{q} + \frac{i}{\hbar} \langle \ldots \mathcal{S}_{\text{nq}}^{(4)} \rangle_\text{q},
\end{equation}
where the first term is the quadratic contribution and the second originates from $\mathcal{S}_{\text{nq}}^{(4)}$. The averages $\langle \ldots \rangle_{\text{q}} $ are taken with respect to the quadratic action $\mathcal{S}_{\text{q}}$ defined in Eq.~\eqref{Seff} at weak transmission.

In the Keldysh field theory formalism the number of photons is defined on the $\pm$ contour by $\langle n \rangle = \langle a_{-}^{*}a_{+} \rangle$. 
Therefore, the non-quadratic contribution, $\langle n \rangle_{\text{nq}} = i\langle a_{-}^{*}a_{+}\mathcal{S}_{\text{nq}}^{(4)} \rangle/\hbar$, 
is
\begin{align} \label{npng0}
\langle n \rangle_{\text{nq}} &= \frac{i}{\hbar} C \int dt_{1}dt_{2}d\epsilon_{l}d\epsilon_{r}\cos[eV(t_{1}-t_{2})/\hbar]e^{-i\omega_{lr}(t_{1}-t_{2})} \nonumber \\
&\times f(\epsilon_{r})\bar{f}(\epsilon_{l})(2\langle a_{-}^{*}a_{+}[\varphi_{+}(t_{2})-\varphi_{-}(t_{1})]^{4}\rangle_\text{q} \nonumber \\
&-\mathsmaller{\sum}\limits_{\sigma}\langle a_{-}^{*}a_{+}[\varphi_{\sigma}(t_{2})-\varphi_{\sigma}(t_{1})]^{4}\rangle_\text{q}).
\end{align}
This expectation value is evaluated using Wick's theorem. Details of this calculation are presented in the Appendix \ref{appB}. The number of 
photons is 
\begin{align} \label{npng}
\langle n \rangle &= \frac{\bar{S}(-\omega_{0})}{2\hbar\omega_{0}}-\frac{\lambda^{2}}{8(\hbar\omega_{0})^{3}}\left[\bar{S}^{2}(-\omega_{0})
\bar{S}(2\omega_{0})\right. \nonumber \\
&\left. -\bar{S}^{2}(\omega_{0})\bar{S}(-2\omega_{0}) \right].
\end{align}
The first term is $\langle n \rangle_\text{q}$ given by Eq.~\eqref{nq}. The second and smaller term originates from the non-quadratic part of the action and describe departure from the thermal-like state. It can be given a physically more transparent form by using $\langle n \rangle_\text{q}=\bar{S}(-\omega_0)/2\hbar\omega_0$ and $\langle n \rangle_\text{q}+1=\bar{S}(\omega_0)/2\hbar\omega_0$ such that the second term of Eq.~\eqref{npng} reads
\begin{align}\label{decomp}
&\frac{\bar{S}^{2}(-\omega_{0})}{(2\hbar\omega_{0})^{2}}\bar{S}(2\omega_{0})-\frac{\bar{S}^{2}(\omega_{0})}{(2\hbar\omega_{0})^{2}}\bar{S}(-2\omega_{0})\nonumber \\
&=\langle n \rangle_\text{q}^2\bar{S}(2\omega_{0})-(\langle n \rangle_\text{q}+1)^2\bar{S}(-2\omega_{0})
\end{align}
To interpret this decomposition, we have to recall that $\bar{S} (\pm \omega)$ is proportional to the absorption/emission noise for the QPC. 
The first term in Eq.~\eqref{decomp} therefore corresponds to the absorption of a pair of photons from the cavity conditioned by their 
presence $\propto \langle n \rangle_\text{q}^2$, while the second term describes photon-pair stimulated emission 
$\propto (\langle n \rangle_\text{q}+1)^2$.
\begin{figure}[!h]
\begin{center}
\includegraphics[width=0.48\textwidth]{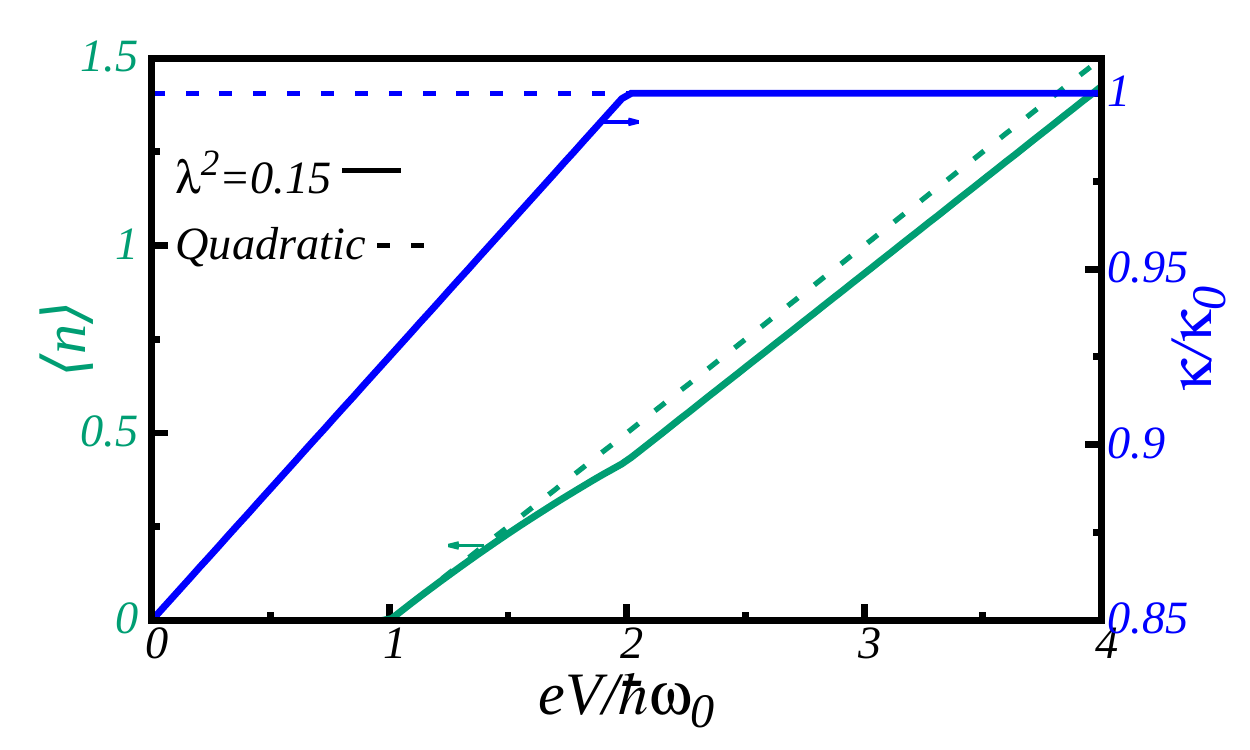}
\caption{Number of photons (left) and cavity damping (right) as a function of the voltage $V$ for $\lambda^2=0.15$ (solid line) and their 
quadratic 
contribution (dash line).\label{fig2}}
\end{center}
\end{figure}

The balance between these two-photon processes depends crucially on the dc-bias. In the range $\hbar \omega_{0}/e < V < 2\hbar \omega_{0}/e$, only 
two-photon absorption occurs since $\bar{S} (-2\omega_0) = 0$, meaning that the available energy from tunneling electrons is not sufficient to excite 
pairs of photons with energy $2 \hbar \omega_0$. As a result, the number of cavity photons~\eqref{npng} is smaller than the quadratic prediction. 
This is shown Fig.~\ref{fig2} where the dependence on the dc-bias is illustrated for a moderate e-p coupling $\lambda^2 = 0.15$. 

For $V>2\hbar\omega_0/e$ the emission of pairs of photons to the cavity is allowed but two-photon absorption still dominates since
\begin{equation}
\bar{S}^{2}(-\omega_{0})\bar{S}(2\omega_{0})-\bar{S}^{2}(\omega_{0})\bar{S}(-2\omega_{0}) = 4(\hbar\omega_{0})^{3}. 
\end{equation} 
Using that $\lambda^{2}\hbar\omega_{0} = E_c$, as noticed in Sec.~\ref{qpc-tl}, we rewrite the mean number of cavity photons as 
\begin{equation} \label{npV2}
\langle n \rangle = \frac{eV-\hbar\omega_{0}-E_{c}}{2\hbar \omega_{0}} \qquad \text{for} \qquad V>2\hbar\omega_0/e,
\end{equation}
in which the electromagnetic charging energy $E_{c}$ of the cavity is subtracted to the available energy $e V$ of electrons. This result is strongly 
reminiscent of the DCB effect at large voltage - or even the genuine Coulomb blockade - in the $P(E)$ 
framework~\cite{ingold-nazarov-book1992}. We elaborate later on this idea in Sec.~\ref{sec4} by establishing an out-of-equilibrium $P(E)$ approach for 
this model and by showing that the reduction of $e V$ by $E_{c}$ is interpreted as a backaction effect of the state of the cavity.

The cavity damping rate is obtained from the retarded self-energy computed with the correction $\mathcal{S}_{\text{nq}}^{(4)}$. The result assumes the form
\begin{align}\label{damping1}
\kappa &= \kappa_0 \left[1- \frac{\lambda^{2}}{2\hbar\omega_0}(\bar{S}(\omega_0)+\bar{S}(-\omega_0))\right] + \frac{\lambda^4}{e^2} S_{rr}(0) \nonumber \\
&+\frac{\lambda^{4}}{2\hbar\omega_0 e^2}[S_{rr}(-\omega_0)\bar{S}(2\omega_0)-S_{rr}(\omega_0)\bar{S}(-2\omega_0)],
\end{align}
where the last term recovers the competition between two-photon emission and absorption. The net effect of this competition is to increase $\kappa$ as two-photon absorption dominates over emission. In addition, there is a Franck-Condon reduction of the leading damping rate $\kappa_0$. The second term $\propto S_{rr} (0)$ describes a process 
in which a single photon is absorbed and reemitted with no energy cost for electrons. Since absorption is $\propto \langle n \rangle_\text{q}$ and emission $\propto \langle n \rangle_\text{q}+1$, the net effect is positive for the damping rate.

The bias voltage dependence of the damping rate is shown in Fig.~\ref{fig2} for $\lambda^2 = 0.15$. Interestingly, the different non-quadratic corrections compensate each other for $e V > 2 \hbar \omega_0$ and we obtain $\kappa = \kappa_0$ in this case. In the range $0 \leq V \leq 2 \hbar\omega_{0}/e$, $\kappa$ is smaller than $\kappa_0$ but increases linearly with the dc-bias. For $e V \leq \hbar \omega_0$, the cavity is in the vacuum state. The expression of the damping rate simplifies as
\begin{equation}
\kappa = \frac{\lambda^2}{e^2} \left[(1-\lambda^2)S_{rr}(\omega_0) + \lambda^{2}S_{rr}(0)\right],
\end{equation}
showing that the absorption of photons with energy $\hbar\omega_0$ is reduced by the Franck-Condon factor $(1-\lambda^2/2)^2$ affecting the electronic transmissions. The linear voltage dependence comes here from the zero-energy photon emission/absorption $\propto S_{rr}(0)$.

We finally compute $g^{(2)}(0)=\langle a_{-}^{* 2}a_{+}^{2}\rangle/\langle n \rangle^{2}$. Using Eq.~\eqref{eqaverage} we write its numerator as
\begin{align}
\langle a_{-}^{* 2}a_{+}^{2}\rangle &= 2\langle n \rangle^{2}+\frac{i}{\hbar}\langle a_{-}^{* 2}a_{+}^{2}\mathcal{S}_{\text{nq}}^{(4)}\rangle_{\text{q,fc}}.
\end{align}
The first term comes from pairing each $a^*_-$ with a $a_+$ field, $\mathcal{S}_{\text{nq}}^{(4)}$ being contracted with one pair only. 
The second term is averaged with the quadratic part of the action $\mathcal{S}_{\text{q}}$ and only the contraction pairing each field 
$a^*_-$ or $a_+$ to a field from $\mathcal{S}_{\text{nq}}^{(4)}$ is kept (fully connected diagram), with the result 
\begin{align}
\frac{i}{\hbar} & \langle a_{-}^{* 2}a_{+}^{2}\mathcal{S}_{\text{nq}}^{(4)}\rangle_{\text{q,fc}}=-\frac{\lambda^{2}}{32(\hbar\omega_{0}^{4})}[\bar{S}(\omega_{0})+\bar{S}(-\omega_{0})]\nonumber \\
&\times [\bar{S}^{2}(-\omega_{0})\bar{S}(2\omega_{0})-\bar{S}^{2}(\omega_{0})\bar{S}(-2\omega_{0})].
\end{align}
Combining this expression with Eq.~\eqref{npng} for the number of photons, we find
\begin{align} \label{g2-final}
g^{(2)}(0)=2-\frac{\lambda^{2}}{8(\hbar\omega_{0})^{2}\bar{S}^{2}(-\omega_{0})}[\bar{S}(\omega_{0})+\bar{S}(-\omega_{0})] \nonumber \\
\times [\bar{S}^{2}(-\omega_{0})\bar{S}(2\omega_{0})-\bar{S}^{2}(\omega_{0})\bar{S}(-2\omega_{0})].
\end{align}
The deviation from the quadratic prediction $g^{(2)}(0)=2$ involves again a balance between two-photon emission and absorption.
\begin{figure}[!h]
\begin{center}
\includegraphics[width=0.5\textwidth]{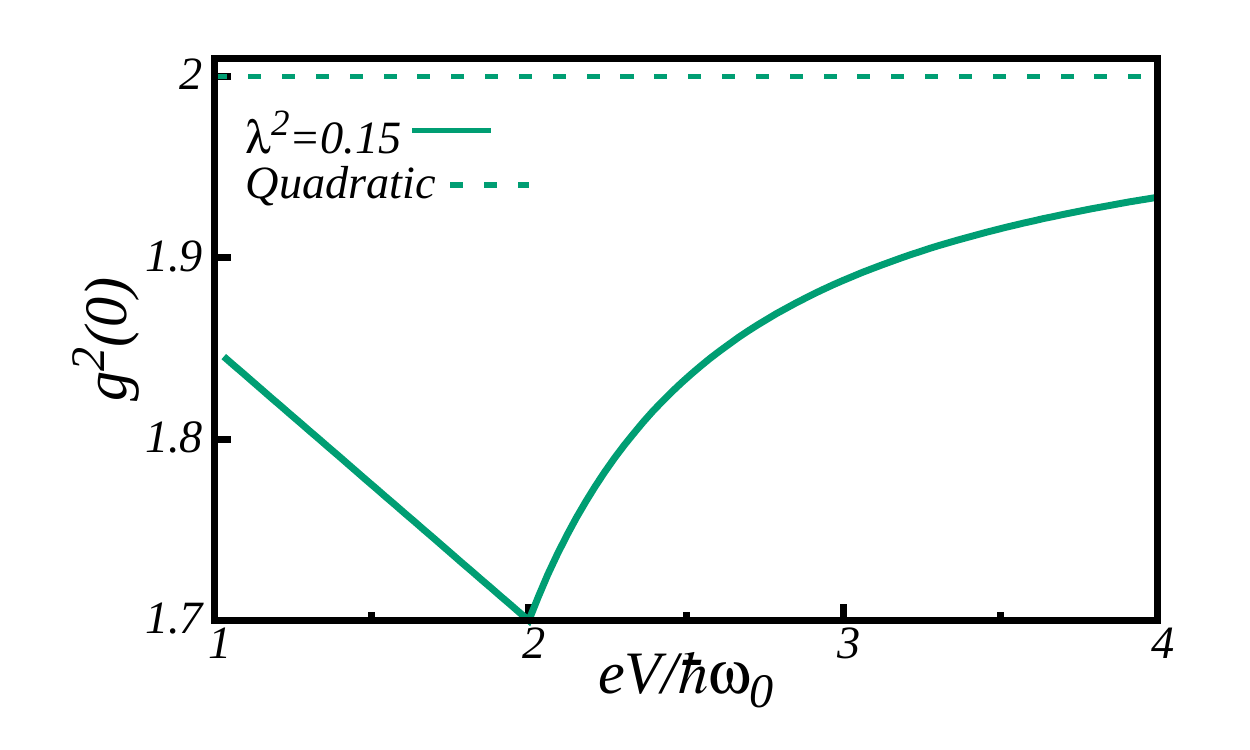}
\caption{$g^{(2)}(0)$ as a function of the voltage $V$ for $\lambda^2=0$ (dash line) and $\lambda^2=0.15$ (solid line).\label{fig-g2}}
\end{center}
\end{figure}
For a low number of photons~\cite{kubala2015}, $g^{(2)}(0) = 2 P_2/P_1^2$, where $P_n$ is the probability to host $n$ photons in the cavity. Two-photon absorption reduces $P_2$ in comparison to $P_1^2$. Since it is more efficient than two-photon emission, $g^{(2)}(0)$ is found to be smaller than $2$ for all voltages. This is shown in Fig.~\ref{fig-g2}. For $e V < \hbar \omega_0$, no photon are present and $g^{(2)}$ vanishes. In the range $\hbar\omega_0 < e V \leq 2\hbar\omega_{0}$, two-photon emission is prohibited, Eq.~\eqref{g2-final} 
simplifies to
\begin{align}
g^{(2)}(0)&=2-\lambda^2(2\langle n \rangle_\text{q} + 1) \nonumber \\
&=2-\frac{\lambda^2}{(2\hbar\omega_{0})^{2}}eV\bar{S}(2\omega_{0}), 
\end{align} 
and $g^{(2)}(0)$ decreases with voltage as the cavity population increases and two-photon absorption processes $\langle n \rangle_\text{q}^2$ are reinforced. For a voltage larger than $2 \hbar \omega_0/e$, two-photon emission sets in and Eq.~\eqref{g2-final} takes the form
\begin{equation}
g^{(2)}(0)=2-E_c\frac{eV}{(eV-\hbar\omega_{0})^{2}},
\end{equation}
increasing with the voltage. At large voltage $e V \gg \hbar \omega_0, E_c$, the quadratic result $g^{(2)}(0)=2$ is finally recovered.

\section{Rate equations \label{sec3}}

The RWA used so far averages to zero terms that do not conserve energy. It removes most off-diagonal elements of 
the density matrix. In this section, we apply a rate equation approach to the QPC-TLC system, corresponding to a quantum master equation 
approach in which off-diagonal elements are disregarded, and indeed recover most results from the previous section. In this way the physical 
picture of two-photon processes is further justified.

\begin{figure}[!h]
\begin{center}
\includegraphics[width=0.48\textwidth]{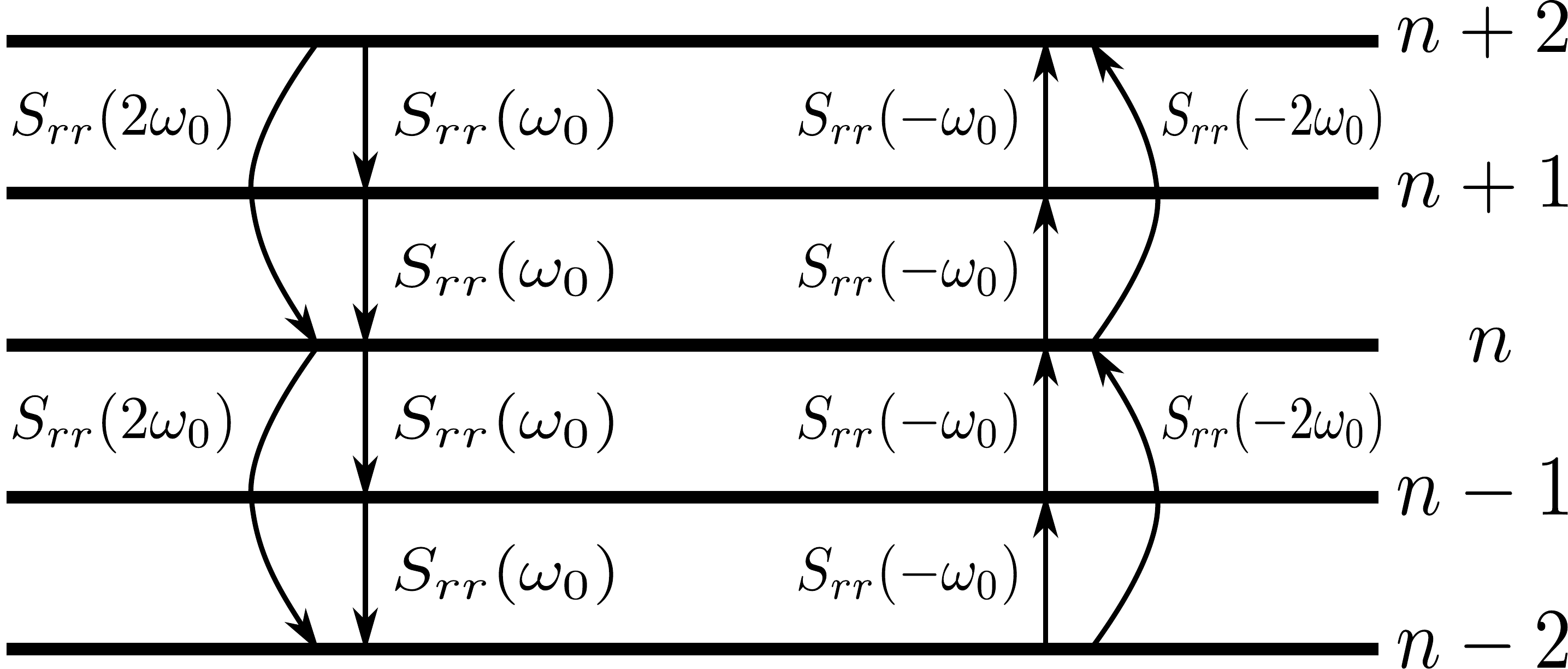}
\caption{Schematic representation of all allowed emission/absorption processes. The cavity emission and absorption processes are defined 
by the QPC absorption [$S_{rr}(\omega)$] and emission [$S_{rr}(-\omega)$] noise. \label{fig-trans}}
\end{center}
\end{figure}

$P_n$ is the probability to have $n$ photons in the cavity. Its time evolution is fixed by
\begin{align} \label{rateq0}
\dot{P}_n &= -(\Gamma_{n\rightarrow n+1}+\Gamma_{n\rightarrow n-1}+\Gamma_{n\rightarrow n+2}+\Gamma_{n\rightarrow n-2})P_n \nonumber \\
&+\Gamma_{n+1\rightarrow n}P_{n+1}+\Gamma_{n-1\rightarrow n}P_{n-1}+\Gamma_{n+2\rightarrow n}P_{n+2} \nonumber \\
&+\Gamma_{n-2\rightarrow n}P_{n-2},
\end{align}
where $\Gamma_{i\rightarrow j}$ denotes the rate from $|i\rangle$ to $|j\rangle$ photons. $\Gamma_{n\rightarrow n\pm1}$ corresponds to 
single-photon and $\Gamma_{n\rightarrow n\pm2}$ to two-photon emission/absorption. These rates are calculated in Appendix~\ref{appC} 
via Fermi-Golden rule in the limit of weak transmissions $T_n \ll 1$. Fig.~\ref{fig-trans} illustrates the ladder of transitions processes. 

The leading (second in $\lambda$) order in the e-p coupling involves only single-photon exchange. Setting $\Gamma_{n\rightarrow n\pm2}=0$, 
the steady state solution of Eq.~\eqref{rateq0} is the Bose-Einstein distribution
\begin{equation}\label{bose-einstein}
P_{n}^{0}=\left(1-\frac{S_{rr}(-\omega_0)}{S_{rr}(\omega_0)}\right)\left(\frac{S_{rr}(-\omega_0)}{S_{rr}(\omega_0)}\right)^n = \frac{\langle n \rangle_\text{q}^n }{(\langle n \rangle_\text{q}+1)^{n+1} },
\end{equation}
corresponding to a thermal Gaussian state. The mean number of photons $\sum_n n P_{n}^{0}$ recovers $\langle n \rangle_\text{q}$ given 
in Eq.~\eqref{nq}.

Two-photon rates are higher orders in $\lambda$ and can be treated perturbatively with respect to the distribution $P_{n}^{0}$. Writing $P_{n}=P_{n}^{0}+p_{n}$, we look for the steady-state solution $\dot{P}_n = 0$ of Eq.~\eqref{rateq0}. Expanding to lowest non-vanishing 
order in $\lambda$ gives
\begin{align} \label{eqrat1}
&-[(n+1)S_{rr}(-\omega_0)+n S_{rr}(\omega_0)]p_{n} +(n+1)S_{rr}(\omega_0)p_{n+1}  \nonumber \\[2mm]
&+n S_{rr}(-\omega_0)p_{n-1}=  \frac{\lambda^2}{4} \Big( -(n+1)(n+2) S_{rr}(2 \omega_0)P_{n+2}^{0} \nonumber \\[2mm]
&+[(n+1)(n+2) S_{rr}(-2 \omega_0)+n(n-1) S_{rr}(2 \omega_0)]P_{n}^{0} \nonumber \\[2mm]
&-n(n-1) S_{rr}(-2 \omega_0)P_{n-2}^{0} \Big). 
\end{align}
The correction to the mean number of photon is determined by multiplying this expression by $n$ and summing over $n$. We finally obtain 
that $\langle n \rangle = \sum_n n (P_{n}^{0}+p_{n})$, calculated perturbatively, coincides with Eq.~\eqref{npng} from Sec.~\ref{sec2}.

We proceed in two steps in order to compute $g^{(2)}(0)$. We multiply Eq.~\eqref{eqrat1} by $n^2$ and sum over $n$ to obtain
\begin{align}
\sum_n & n (n-1) p_{n}  =-\frac{\lambda^{2}}{32(\hbar\omega_{0})^{4}}[9\bar{S}(-\omega_{0}) + \bar{S}(\omega_{0})] \nonumber \\
&\times [\bar{S}^{2}(-\omega_{0})\bar{S}(2\omega_{0})-\bar{S}^{2}(\omega_{0})\bar{S}(-2\omega_{0})]
\end{align}
corresponding to the two-photon correction to the average $\langle a^{\dagger 2} a^{2} \rangle$ while the leading order is simply
$\sum_n n (n-1) P_{n}^{0} = 2 (\sum_n n P_{n}^{0})^2 = 2 \langle n \rangle_\text{q}^2$. We then include the denominator $\langle n \rangle^2$ 
expanded in $\lambda$ and recover exactly Eq.~\eqref{g2-final} from Sec.~\ref{sec2}.

These results not only reinforces our physical interpretation for the formulas derived in Sec.~\ref{sec2} but also shows that the cavity 
properties in the weak coupling regime are well described by a diagonal density matrix. 

\section{Dynamical backaction \label{sec4}}

The QPC-TLC hybrid system (with weak transmissions) can also be discussed within an enlightening approach which emphasizes backaction. 
The cavity provides a readout of the noise power spectrum of the QPC or tunnel junction. In return, there is backaction from the cavity 
with a DCB effect which reduces transport. The modified noise properties of electrons are imprinted 
in the state of the cavity. This effect is captured by the fourth order in $\lambda$ calculation of the previous sections but it can be 
made more explicit by extending the $P(E)$ theory to a non-equilibrium steady state situation.

We begin by assuming that the cavity is in the thermal state characterized by the photon distribution~\eqref{bose-einstein} and the mean 
number of photons is $\langle n \rangle_\text{q}$, see Eq.~\eqref{nq}. For weak transmissions $T_n \ll 1$, we use the tunnel Hamiltonian $\bar{H}_{T}=\hat{\mathcal{T}}e^{-\lambda(\hat{a}^{\dagger}-\hat{a})}+\hat{\mathcal{T}}^{\dagger}e^{\lambda(\hat{a}^{\dagger}-\hat{a})}$, 
where the cavity field $\hat{a}$ plays the role of the environment, and proceed with the $P(E)$ approach by computing the current noise 
correlator. The tunneling limit allows for a factorization of the electron and environment variables~\cite{safi2014,altimiras2014} such 
that the noise takes the convoluted form
\begin{equation}
S_\text{DCB}(\omega) = \int_{-\infty}^{\infty}S_{rr} (\hbar\omega-E)P(E)dE.
\end{equation}
$S_{rr}$ ($S_\text{DCB}$) is the noise in the absence (presence) of the cavity. The $P(E)$ function,
\begin{equation}
P(E) = \frac{1}{2\pi\hbar}\int dt e^{iEt/\hbar}\langle \hat{X}^{\dagger}(t)\hat{X}(0)\rangle,
\end{equation}
where $\hat{X}(t)=\exp[-\lambda(\hat{a}^{\dagger}(t)-\hat{a}(t))]$ characterizes the environment. For $E>0$, it gives the probability for 
the QPC/tunnel junction to emit a photon of energy $E$ in the environment during a tunneling event. The $E<0$ part describes photon absorption. 

$P(E)$ can be evaluated exactly in a Gaussian state using Wick's theorem. We nevertheless expand in $\lambda$ for consistency with our weak 
coupling scheme, and find
\begin{align}
P(E) &= (1-\lambda^2(\langle n \rangle_\text{q} + n_1))\delta(E) + \lambda^2 \langle n \rangle_\text{q} \delta(E+\hbar\omega_0) \nonumber \\
&+ \lambda^2 n_1 \delta(E-\hbar\omega_0)
\end{align}
where $n_1 = \langle n \rangle_\text{q} + 1$. It can be check that $P(E)$ is normalized, $\int d E P(E) = 1$. This expression is valid at zero temperature. The presence of non-zero values for $E<0$ therefore indicates that the cavity is in an out-of-equilibrium state and is able to 
provide energy to the quantum conductor. For $e V < \hbar \omega_0$, the cavity is empty and $\langle n \rangle_\text{q} = 0, n_1 = 1$; 
the elastic peak $\delta (E)$ is renormalized by vacuum fluctuations~\cite{jin-schon-prb2015} and inelastic photon emission occurs for 
$E = \hbar \omega_0$.

We consider the noise properties of the QPC. The absorption noise is
\begin{align} \label{Anoise}
S_{\text{DCB}}(\omega_0)&=(1-\lambda^2(\langle n \rangle_\text{q}+n_1))S_{rr}(\omega_0)+ \lambda^2 n_1 S_{rr}(0) \nonumber \\
&+ \lambda^2 \langle n \rangle_\text{q} S_{rr}(2\omega_0).
\end{align}
In addition to the renormalized single-photon absorption, the second term describes the correlated emission and reabsorption of a photon by the QPC, present also when the cavity is in a vacuum state, and the third term, pair absorption, requires an occupied cavity.

A similar analysis for the emission noise 
\begin{align} \label{Enoise}
S_\text{DCB}(-\omega_0)&=(1-\lambda^2(\langle n \rangle_\text{q}+n_1))S_{rr}(-\omega_0) \nonumber \\
&+ \lambda^2 \langle n \rangle_\text{q} S_{rr}(0)+ \lambda^2 n_1 S_{rr}(-2\omega_0)
\end{align}
reveals a renormalized single-photon emission, a correlated photon absorption and reemission, and pair emission. The first and second 
 terms are non-zero only for $e V > \hbar \omega_0$, the third term for $e V > 2 \hbar \omega_0$.

To summarize, the cavity back-action provides new absorption and emission mechanisms. Their effect on the noise properties of a tunnel 
junction have been studied in Ref.~\onlinecite{jin-schon-prb2015}. We focus here on the effect on the cavity state. To lowest order in 
the e-p coupling, the damping rate and number of photons of the cavity 
\begin{subequations}
\begin{align}
\kappa_\text{DCB} &= \frac{\lambda^2}{e^2} [S_{\text{DCB}}(\omega_0)-S_{\text{DCB}}(-\omega_0)]  \\[2mm]
\langle n \rangle_\text{DCB} &= \frac{\lambda^2 S_{\text{DCB}}(-\omega_0)}{e^2 \kappa_\text{DCB}}.
\end{align} 
\end{subequations}
can be computed using Eqs.~\eqref{Anoise} and \eqref{Enoise} and the relations $\langle n \rangle_\text{q}=\bar{S}(-\omega_0)/2\hbar\omega_0$ 
and $n_1=\bar{S}(\omega_0)/2\hbar\omega_0$. We obtain 
\begin{align}
&\kappa_\text{DCB} = \kappa_0 \left[1- \frac{\lambda^{2}}{2\hbar\omega_0}(\bar{S}(\omega_0)+\bar{S}(-\omega_0))\right] + \frac{\lambda^4}{e^2} S_{rr}(0) \nonumber \\
&+\frac{\lambda^{4}}{2\hbar\omega_0 e^2}[S_{rr}(-\omega_0)\bar{S}(2\omega_0)-S_{rr}(\omega_0)\bar{S}(-2\omega_0)]\\
&\langle n \rangle_\text{DCB} = \frac{\bar{S}(-\omega_{0})}{2\hbar\omega_{0}}-\frac{\lambda^{2}}{8(\hbar\omega_{0})^{3}}\left[\bar{S}^{2}(-\omega_{0})
\bar{S}(2\omega_{0})\right. \nonumber \\
&\left. -\bar{S}^{2}(\omega_{0})\bar{S}(-2\omega_{0}) \right].
\end{align} 
These expressions coincide exactly with those obtained in Sec.~\ref{nq-action}, see Eqs.~\eqref{damping1} and~\eqref{npng}.

\section{Summary and conclusion \label{concl}}

We investigated the properties of a single-mode cavity field coupled to a quantum point contact or tunnel junction. We first used a 
Keldysh path integral framework using two formulations related by a unitary gauge transformation: a coupling between the quantum 
voltage operator of the cavity and the lead densities, and an inductive coupling where each scattering event is accompanied by the 
excitation of the cavity. Expanding for weak electron-photon coupling to a quadratic form, we found a Gaussian thermal field 
distribution with $g^{(2)} (0)=2$, and the photon self-energies are fully characterized by the emission and absorption noise of the 
quantum point contact. The damping rate is constant in this limit, independent of the bias voltage.

Proceeding with the next order in the electron-photon coupling, we identified two-photon processes: pair emission or absorption as 
well as correlated photon emission and absorption. We recovered these effects using a rate equation approach and a $P(E)$ calculation 
adapted to our non-equilibrium situation. We obtained a reduced number of photons relatively to the thermal state prediction, a 
suppressed $g^{(2)} (0) < 2$ for a not too large bias voltage, and a reduced cavity damping rate for $e V< 2 \hbar \omega_0$, which are 
explained by noting the preeminence of two-photon absorption over emission.

It is tempting to speculate about the extension of these findings to higher orders in the electron-photon coupling $\lambda$. Increasing 
powers of $\lambda$ will introduce emission and absorption processes involving three, four and higher number of photons, in which photon 
absorption is always energetically more favourable than emission. This should modify the photon distribution by adding more weight to small 
number of photons compared to larger numbers as the numerical results of Ref.~\onlinecite{BergenfeldtSET2012} seem also to indicate. 
Interestingly, the balance between emission and absorption depends on the bias voltage such that high voltages are expected to bring the 
system back to a Gaussian thermal distribution.

We emphasize a strong difference with dc-biased Josephson junctions where the physics of strong electron-photon coupling is essentially 
described by generalized Frank-Condon factors and only processes conserving energy occur as there is no electronic dissipation. In tunnel 
junctions, the non-conserved part of the energy can be dissipated in the leads which allows for more processes. The question is still 
open whether it is possible to realize a non-classical state with $g^{(2)} (0) <1$ apart from the region $e V \simeq \hbar \omega_0$. 
As a prospect of future investigation, we mention the extension of our approach by including fourth-order current correlations. They 
appear to the order considered in this work when the QPC transmission probabilities are no longer small.

\section{Acknowledgments} We thank C. Altimiras, A. Clerk, P. Joyez, F. Portier and P. Simon for fruitful discussions. U.C.M. acknowledges the support from CNPq-Brazil 
(Project No. 229659/2013-6).
\appendix

\section{Alternative gauge formulation \label{appA}}

In this appendix we detail the Keldysh path integral framework presented in Sec.~\ref{sec2}.

\subsection{Quadratic effective action}

We discuss the equivalence between the Keldysh path integral formulations presented in Sec.~\ref{sec1} and \ref{sec2}. 
To show the correspondence we rederive the quadratic action Eq.~\eqref{Seff} starting form the action
\begin{equation} \label{eq-epaction}
\mathcal{S}_{\text{e-p}} = -\frac{i\hbar}{2}\sum_{n} \Tr \ln \left[1 + \frac{T_{n}}{4}\left(\{ \check{G}_l,\check{G}_r\}-2\right)\right].
\end{equation}
Using $\check{G}_l(t,t^\prime) = \check{U}^\dagger(t) \check{G}_\text{eq}^l(t-t^\prime) \check{U}(t^\prime)$ and $\check{G}_r(t,t^\prime)=\check{G}_\text{eq}^r(t-t^\prime)$ 
($\check{G}_\text{eq}^\alpha$ and $\check{U}(t^\prime)$ are defined in Eqs.~\eqref{eqKGF} and \eqref{gauge-eq}), the anticommutator can 
be written as
\begin{equation}
\{\check{G}_l(t,t^\prime),\check{G}_r(t^\prime,t)\} = 2 + 2\check{M}(t,t^\prime),
\end{equation}
where 
\begin{equation} \label{matrixM}
\check{M}(t,t^\prime) = \int \frac{d\epsilon_l d\epsilon_r}{(2\pi \hbar)^2} e^{-i\omega_{lr}(t-t^\prime)} 
\begin{pmatrix}
 m_1(\mathbf{x}) &  m_2(\mathbf{x}) \\ m_3(\mathbf{x}) & m_4(\mathbf{x}) 
\end{pmatrix},
\end{equation}
$\omega_{lr}=(\epsilon_l - \epsilon_r)/\hbar$, $\mathbf{x}=\{t,t^\prime,\epsilon_l,\epsilon_r\}$ and 
\begin{subequations} \label{melem}
\begin{align}
&m_1(\mathbf{x}) = 2f(\epsilon_l)\bar{f}(\epsilon_r)[e^{\lambda(\varphi_{-}(t^\prime)-\varphi_{+}(t))}-e^{\lambda(\varphi_{+}(t^\prime)-\varphi_{+}(t))}] \nonumber \\
&+2f(\epsilon_r)\bar{f}(\epsilon_l)[e^{\lambda(\varphi_{+}(t^\prime)-\varphi_{-}(t))}-e^{\lambda(\varphi_{+}(t^\prime)-\varphi_{+}(t))}] \\
&m_2(\mathbf{x}) = f(\epsilon_r)[1-2f(\epsilon_l)] \sum_{\sigma=\pm}\sigma e^{\lambda(\varphi_{\sigma}(t^\prime)-\varphi_{\sigma}(t))} \\
&m_3(\mathbf{x}) = \bar{f}(\epsilon_r)[1-2f(\epsilon_l)] \sum_{\sigma=\pm}\sigma e^{\lambda(\varphi_{\sigma}(t^\prime)-\varphi_{\sigma}(t))} \\
&m_4(\mathbf{x}) = 2f(\epsilon_l)\bar{f}(\epsilon_r)[e^{\lambda(\varphi_{-}(t^\prime)-\varphi_{+}(t))}-e^{\lambda(\varphi_{-}(t^\prime)-\varphi_{-}(t))}]\nonumber \\
&+2f(\epsilon_r)\bar{f}(\epsilon_l)[e^{\lambda(\varphi_{+}(t^\prime)-\varphi_{-}(t))}-e^{\lambda(\varphi_{-}(t^\prime)-\varphi_{-}(t))}], 
\end{align}
\end{subequations}
where $f(\epsilon)$ is the Fermi function, $\bar{f}(\epsilon)=1-f(\epsilon)$ and $\varphi_{\pm}=a_{\pm}^{*} - a_{\pm}$ is the photon complex 
field defined in the $\pm$ Keldysh time-ordered contour. We use the following expansion strategy: (i) one notices that the matrix $\check{M}$
vanishes for $\lambda=0$ and is linear in $\lambda$ at small coupling, (ii) therefore expanding in $T_n \check{M}$ is consistent with 
an expansion in $\lambda$, (iii) the second order in $T_n \check{M}$,
\begin{equation} \label{eq-ac0}
\mathcal{S}_\text{e-p} = \frac{i\hbar}{4}\sum_n T_n \Tr[\check{M}] + \frac{i\hbar}{8}\sum_n T_n^2 \Tr[\check{M}^2],
\end{equation}
thus exhausts all possible contributions up to second order in $\lambda$. Orders in $T_n$ higher than two must all vanish if we restrict
the action to second order in $\lambda$. Expanding directly the action Eq.~\eqref{eq-epaction} up to $T_n^2$ gives
\begin{align}\label{eq0}
&\mathcal{S}_\text{e-p} =-\frac{i\hbar}{8}\sum_{n}T_{n}\left(1+\frac{T_{n}}{2} \right) \Tr[\{ \check{G}_l,\check{G}_r\}]  \\
&+\frac{i\hbar}{32}\sum_{n}T_{n}^{2}\left(\Tr[\check{G}_l\check{G}_r\check{G}_l\check{G}_r]+\Tr[\check{G}_l\check{G}_r\check{G}_r\check{G}_l]\right). \nonumber
\end{align}
Using 
\begin{equation}
\int dt_1 \check{G}_\text{eq}^{\alpha}(t-t_1)\check{G}_\text{eq}^{\alpha}(t_1-t^\prime) = \check{\tau}_{0}\delta(t-t^\prime),
\end{equation}
where $\tau_0$ is the identity matrix, it can be shown that the third term of Eq.~\eqref{eq0} is equal to the identity. Thus, 
we define the quadratic e-p action as $\mathcal{S}_\text{e-p}=\mathcal{S}_1 +\mathcal{S}_2$, where 
\begin{subequations}
\begin{align}
\mathcal{S}_1 &=-\frac{i\hbar}{8}\sum_{n}T_{n}\left(1+\frac{T_{n}}{2} \right) \Tr[\{\check{G}_l,\check{G}_r\}] \\
\mathcal{S}_2 &=\frac{i\hbar}{32}\sum_{n}T_{n}^{2}\Tr[\check{G}_l\check{G}_r\check{G}_l\check{G}_r].
\end{align}
\end{subequations}

Before proceeding with the derivation of $\mathcal{S}_\text{e-p}$ we rotate the equilibrium GF and $\check{U}$ matrices. 
The rotating matrix is
\begin{equation}
\check{L}= \frac{1}{\sqrt{2}}\begin{pmatrix}
1 && 1 \\ 1 && -1
\end{pmatrix} = L^{\dagger}.
\end{equation}
The rotated equilibrium matrix, defined as $\check{G}_\text{r}^{\rm \alpha} = \check{L}^{\dagger}\check{G}_\text{eq}^{\alpha}\check{L}$, is 
\begin{equation} \label{GF-rot}
\check{G}_\text{r}^{\rm \alpha}(\epsilon_\alpha)=\begin{pmatrix} 1 & 2[1-2f(\epsilon_\alpha)] \\ 0 & -1  \end{pmatrix},
\end{equation}
and the rotated $\check{U}_\text{r}=\check{L}^{\dagger}\check{U}\check{L}$ is conveniently written as
\begin{align*}
\check{U}_\text{r}(t)=A(t)\check{\tau}_{0} + B(t)\check{\tau}_{x},
\end{align*}
where $\check{\tau}_{x}$ is the first Pauli matrix, and 
\begin{subequations}\label{eq-AB}
\begin{align}
A(t) &= \frac{1}{2}[e^{\lambda \varphi_{+}(t)}+e^{\lambda \varphi_{-}(t)}] \nonumber \\
&\approx 1+\frac{\lambda}{\sqrt{2}} \varphi_{c}(t)+\frac{\lambda^{2}}{4}[\varphi_{c}^{2}(t)+\varphi_{q}^{2}(t)] \\
B(t) &= \frac{1}{2}[e^{\lambda \varphi_{+}(t)}-e^{\lambda \varphi_{-}(t)}]  \nonumber \\
&\approx \frac{\lambda}{\sqrt{2}} \varphi_{q}(t)[1+\frac{\lambda}{\sqrt{2}} \varphi_{c}(t)],
\end{align}
\end{subequations}
the above equations were obtained by expanding the exponentials to second-order in $\lambda$ and performing the Keldysh rotation on 
the photonic complex fields $\varphi_\pm = (\varphi_c \pm \varphi_q)/\sqrt{2}$, with $\varphi_{c(q)}=a_{c(q)}^{*}-a_{c(q)}$.

Using the above relations and Fourier transforming the equilibrium GF [$\check{G}_\text{r}^\alpha(t) =\frac{1}{2\pi\hbar} \int d\epsilon_\alpha \check{G}_\text{r}^\alpha(\epsilon_\alpha)e^{-i\epsilon_\alpha t/\hbar}$], we rewrite
\begin{align} \label{aS1}
&\mathcal{S}_1=g_{1} \int dt_{1}dt_{2}d\epsilon_{l}d\epsilon_{r}e^{-i(\omega_{lr} + eV/\hbar)(t_{1}-t_{2})} \left( A^{\dagger}(t_{1})A(t_{2})
 \right. \nonumber \\
&\left. \times\Tr[\check{G}_\text{r}^l(\epsilon_{l})\check{G}_\text{r}^r(\epsilon_{r})] +A^{\dagger}(t_{1})B(t_{2})\Tr[\check{G}_\text{r}^l(\epsilon_{l})\check{\tau}_{x}\check{G}_\text{r}^r(\epsilon_{r})]\right. \nonumber \\
&\left. +B^{\dagger}(t_{1})A(t_{2})\Tr[\check{G}_\text{r}^r(\epsilon_{r})\check{\tau}_{x}\check{G}_\text{r}^l(\epsilon_{l})]+B^{\dagger}(t_{1})B(t_{2})\right. \nonumber \\
&\left. \times \Tr[\check{G}_\text{r}^l(\epsilon_{l})\check{\tau}_{x}\check{G}_\text{r}^r(\epsilon_{r})\check{\tau}_{x}]\right).
\end{align}
where $g_1 = -i\sum_n T_n(1+T_n/2)/16\pi^2\hbar$. To write the above equation we have shifted the energies $\epsilon_{\alpha} \rightarrow \epsilon_{\alpha}+eV_{\alpha}$, $V_{\alpha}$ is the voltage applied to the lead $\alpha$, and defined $V=V_l-V_r$. The trace 
is only over the matrix structure. Analogously, the second term is written as
\begin{widetext}
\begin{align} \label{aS2}
\mathcal{S}_2 &=g_{2} \int d\epsilon_{l}d\epsilon_{r}d\epsilon_{l}^{\prime}d\epsilon_{r}^{\prime} dt_{1}dt_{2}dt_{3}dt_{4}
\left[ A^{\dagger}(t_{1})A(t_{2})A^{\dagger}(t_{3})A(t_{4})\Tr[\check{G}_\text{r}^l(\epsilon_l)\check{G}_\text{r}^r(\epsilon_r) \check{G}_\text{r}^{l^{\prime}}(\epsilon_l^{\prime})\check{G}_\text{r}^{r^{\prime}}(\epsilon_r^{\prime})] \right. \nonumber \\
&\left.+ 2A^{\dagger}(t_{1})A(t_{2})\left(A^{\dagger}(t_{3})B(t_{4})\Tr[\check{G}_\text{r}^l(\epsilon_l)\check{G}_\text{r}^r(\epsilon_r) \check{G}_\text{r}^{l^{\prime}}(\epsilon_l^{\prime})\check{\tau}_{x}\check{G}_\text{r}^{r^{\prime}}(\epsilon_r^{\prime})] + B^{\dagger}(t_{3})A(t_{4})\Tr[\check{G}_\text{r}^l(\epsilon_l)\check{G}_\text{r}^r(\epsilon_r) \check{\tau}_{x} \check{G}_\text{r}^{l^{\prime}}(\epsilon_l^{\prime})\check{G}_\text{r}^{r^{\prime}}(\epsilon_r^{\prime})] \right) \right. \nonumber \\
&\left.+ 2A^{\dagger}(t_{1})B^{\dagger}(t_{3})\left(A(t_{2})B(t_{4})\Tr[\check{G}_\text{r}^l(\epsilon_l)\check{G}_\text{r}^r(\epsilon_r) \check{\tau}_{x}\check{G}_\text{r}^{l^{\prime}}(\epsilon_l^{\prime})\check{\tau}_{x}\check{G}_\text{r}^{r^{\prime}}(\epsilon_r^{\prime})] 
+ B(t_{2})A(t_{4}) \Tr[\check{G}_\text{r}^l(\epsilon_l)\check{\tau}_{x}\check{G}_\text{r}^r(\epsilon_r) \check{\tau}_{x}\check{G}_\text{r}^{l^{\prime}}(\epsilon_l^{\prime})\check{G}_\text{r}^{r^{\prime}}(\epsilon_r^{\prime})]\right)\right. \nonumber \\ 
&\left.+ A^{\dagger}(t_{1})B(t_{2})A^{\dagger}(t_{3})B(t_{4}) \Tr[\check{G}_\text{r}^l(\epsilon_l)\check{\tau}_{x}\check{G}_\text{r}^r(\epsilon_r) \check{G}_\text{r}^{l^{\prime}}(\epsilon_l^{\prime})\check{\tau}_{x}\check{G}_\text{r}^{r^{\prime}}(\epsilon_r^{\prime})]+ B^{\dagger}(t_{1})A(t_{2})B^{\dagger}(t_{3})A(t_{4}) \right. \nonumber \\ 
&\left.  \times \Tr[\check{G}_\text{r}^l(\epsilon_l)\check{G}_\text{r}^r(\epsilon_r) \check{\tau}_{x}\check{G}_\text{r}^{l^{\prime}}(\epsilon_l^{\prime})\check{G}_\text{r}^{r^{\prime}}(\epsilon_r^{\prime})\check{\tau}_{x}]\right] e^{-i(\omega_{lr^{\prime}}+eV/\hbar)t_{1}}e^{-i(\omega_{rl}-eV/\hbar)t_{2}}e^{-i(\omega_{l^{\prime}r}+eV/\hbar)t_{3}}e^{-i(\omega_{r^{\prime}l^{\prime}}-eV/\hbar)t_{4}},
\end{align}
\end{widetext}
where $g_2 = i\sum_n T_n^2/512\pi^4\hbar^3$.

We now focus on $\mathcal{S}_1$. The first step is to compute the traces. Using Eq.~\eqref{GF-rot} we obtain, for instance, 
\begin{align*}
&\Tr[\check{G}_\text{r}^l(\epsilon_{l})\check{G}_\text{r}^r(\epsilon_{r})] =2  \\
&\Tr[\check{G}_\text{r}^l(\epsilon_{l})\check{\tau}_{x}\check{G}_\text{r}^r(\epsilon_{r})]=4[f(\epsilon_r)\bar{f}(\epsilon_l)-f(\epsilon_l)\bar{f}(\epsilon_r)] \\
&\Tr[\check{G}_\text{r}^l(\epsilon_{l})\check{\tau}_{x}\check{G}_\text{r}^r(\epsilon_{r})\check{\tau}_{x}] =2-8[f(\epsilon_r)\bar{f}(\epsilon_l)+f(\epsilon_l)\bar{f}(\epsilon_r)]
\end{align*}
A similar result is obtained for the remaining trace. We rewrite Eq.~\eqref{aS1} as
\begin{align}
&\mathcal{S}_1= 2 g_{1} \int dt_{1}dt_{2}d\epsilon_{l}d\epsilon_{r} e^{-i(\omega_{lr} + eV/\hbar)(t_{1}-t_{2})} \left( A^{\dagger}(t_{1})A(t_{2})  \right. \nonumber \\
&\left.+B^{\dagger}(t_{1})B(t_{2})-4[f(\epsilon_r)\bar{f}(\epsilon_l)+f(\epsilon_l)\bar{f}(\epsilon_r)]B^{\dagger}(t_{1})B(t_{2}) \right. \nonumber \\
&\left. +2[f(\epsilon_r)\bar{f}(\epsilon_l)-f(\epsilon_l)\bar{f}(\epsilon_r)][A^{\dagger}(t_{1})B(t_{2})-B^{\dagger}(t_{1})A(t_{2})]\right)
\end{align}

Using Eqs.~\eqref{eq-AB} one can show that the integrals over the two first terms do not depend on the cavity field variables and can be disregarded. 
The remaining terms are written as
\begin{align}
&\mathcal{S}_1= \frac{4g_{1}\lambda}{\sqrt{2}}  \int dt_{1}dt_{2}d\epsilon_{l}d\epsilon_{r} e^{-i(\omega_{lr} + eV/\hbar)(t_{1}-t_{2})} 
(\varphi_q(t_1)\nonumber \\
&+\varphi_q(t_2))[f(\epsilon_r)\bar{f}(\epsilon_l)-f(\epsilon_l)\bar{f}(\epsilon_r)]\nonumber \\
&+2g_{1}\lambda^2\int dt_{1}dt_{2}d\epsilon_{l}d\epsilon_{r} e^{-i(\omega_{lr} + eV/\hbar)(t_{1}-t_{2})}\{(\varphi_q(t_1)\nonumber \\
&+\varphi_q(t_2))(\varphi_c(t_2)-\varphi_q(t_1))[f(\epsilon_r)\bar{f}(\epsilon_l)-f(\epsilon_l)\bar{f}(\epsilon_r)]\nonumber \\
&+2\varphi_q(t_1)\varphi_q(t_2)[f(\epsilon_r)\bar{f}(\epsilon_l)+f(\epsilon_l)\bar{f}(\epsilon_r)]\}.
\end{align}

Next we Fourier transform $\varphi_{c(q)}(t) = \int_\omega \varphi_{c(q)}(\omega)e^{-i\omega t}$, obtaining
\begin{align} \label{qpc1}
\mathcal{S}_1&=-\frac{i\lambda^{2}}{4\pi}g_0\int_{\omega} (\varphi_{c},\varphi_{q})_{-\omega} 
\begin{pmatrix} 0 & -\hbar\omega \\ \hbar\omega & \bar{Y}(\omega) \end{pmatrix}  \begin{pmatrix} \varphi_{c} \\ \varphi_{q}\end{pmatrix}_{\omega}\nonumber \\
&+i\sqrt{2}\lambda eV g_0\int_{\omega}\varphi_{q}(\omega)\delta(\omega),
\end{align}
where we defined $g_0 = \sum_n T_n(1+T_n/2)$, $\bar{Y}(\omega) = \bar{S}(\omega)+\bar{S}(-\omega)$ and $\bar{S}(\omega)$ is defined 
in Eq.~\eqref{eq-noise}.

Following the same steps we derive the action 
\begin{align} \label{qpc2}
\mathcal{S}_2&=\frac{i\lambda^{2}}{8\pi} \sum_n T_n^2\int_{\omega} (\varphi_{c},\varphi_{q})_{-\omega}  \begin{pmatrix}
0 && -\hbar\omega \\ \hbar\omega && \bar{Y}_1(\omega) \end{pmatrix}  \begin{pmatrix} \varphi_{c} \\ \varphi_{q}\end{pmatrix}_{\omega} \nonumber \\
&-i\frac{\sqrt{2}}{2}\lambda eV \sum_n T_n^2 \int_{\omega} \varphi_{q}(\omega)\delta(\omega),
\end{align}
where 
\begin{equation*}
\bar{Y}_1(\omega) =2[ \bar{Y}(\omega)+\bar{S}(\omega)+\hbar\omega-4\hbar\omega\Theta(\hbar\omega)].
\end{equation*}
Adding Eqs.~\eqref{qpc1} and \eqref{qpc2} and changing variables $\varphi_{c(q)} = a_{c(q)}^{*}-a_{c(q)}$ we obtain the same quadratic e-p action~\eqref{Sep2} 
as in Sec.~\ref{sec1}. The quadratic action~\eqref{Seff} is finally exactly recovered (with the same self-energies) by applying the same RWA as in Sec.~\ref{sec1}.

A difference occurs nevertheless in the linear term in the action. In this case, it assumes the form
\begin{align} \label{eq-slinear}
\mathcal{S}_\text{L}&=i 2\sqrt{2}\pi \hbar \lambda \langle I \rangle  \int_{\omega}[a_{q}^{*}(-\omega)-a_{q}(\omega)]\delta(\omega), 
\end{align}
where $\langle I \rangle = eV\sum_n T_n/h$ is the QPC current, in contrast with the linear term~\eqref{s-linear} obtained in
Sec.~\ref{sec1} for the other gauge. This difference is however expected since a change of gauge, via a unitary transformation, also
modifies observables~\cite{cohen1997}. With the operator $\hat{U}_0$, it merely shifts field operators. The linear term~\eqref{eq-slinear}
can be removed by shifting the classical field
\begin{equation} \label{shifclass}
a_{c}=a_{c}- 2\sqrt{2}\pi \hbar i\lambda \langle I \rangle \delta(\omega)G_{\text{R}}(\omega).
\end{equation}

\subsection{Non-quadratic action}

Here we show how to obtain the non-quadratic actions $\mathcal{S}_\text{nq}^{(3)}$ and $\mathcal{S}_\text{nq}^{(4)}$ used in Sec.~\ref{nq-action}. 
It is convenient to consider the first term of the action defined in Eq.~\eqref{eq-ac0}. Considering $\lambda \ll 1$ we first expand the exponentials 
in the matrix elements [Eqs.~\eqref{melem}] of the matrix $\check{M}$ to fourth-order, then we take the trace over the matrix structure and, finally, 
we shift the energies $\epsilon_\alpha$ by the voltage applied to the lead $\alpha$, i.e, $\epsilon_{l(r)} \rightarrow \epsilon_{l(r)}\pm V/2$. Thus, 
obtaining
\begin{align}
&\mathcal{S}_\text{eff} =-\frac{i g_c}{8\pi^2 \hbar} \sum_{p=1}^{4} \frac{\lambda^p}{p!} \int dt_1 dt_2 d\epsilon_l d\epsilon_r f(\epsilon_r)\bar{f}(\epsilon_l)e^{-i\omega_{lr}(t-t^\prime)} \nonumber \\
&\times [2(\varphi_+(t_2)-\varphi_-(t_1))^{p}-\sum_{\sigma}(\varphi_\sigma(t_2)-\varphi_\sigma(t_1))^{p}] \nonumber \\
&\times \left(e^{-ieV(t_1-t_2)/\hbar}+(-1)^p e^{ieV(t_1-t_2)/\hbar}\right),
\end{align}
where $g_c= \sum_n T_n$. The terms $p=1$ and 2 will produce the linear [Eq. \eqref{eq-slinear}] and quadratic [Eq. \eqref{Sep2}] actions. 
$\mathcal{S}_{\text{nq}}^{(3)}$ [Eq.~\eqref{eq-seff3}] and $\mathcal{S}_{\text{nq}}^{(4)}$ [Eq.~\eqref{eq-seff4}] are given by terms $p=3$ and 4, respectively. 
As they are defined by the difference between the complex fields, $\varphi_{\sigma}-\varphi_{\sigma^\prime}$, the shift of classical fields, Eq.~\eqref{shifclass}, 
does not alter the form of $\mathcal{S}_{\text{nq}}^{(3)}$ and $\mathcal{S}_{\text{nq}}^{(4)}$.

\section{Correction to the number of photons \label{appB}}

We present here the main steps used in the derivation of the non-quadratic contribution to mean number of photons $\langle n \rangle_\text{nq} = \frac{i}{\hbar}\langle a_{-}^{*}a_{+} \mathcal{S}_{\text{nq}}^{(4)} \rangle$. For simplicity we set $\hbar=1$ and $e=1$. Our starting point is 
Eq.~\eqref{npng0}, we rewrite it as
\begin{align} \label{np1}
\langle n \rangle_\text{nq} &= C_1 \sum_{\beta=\pm} \int dt_{1}dt_{2}d\epsilon_{l}d\epsilon_{r}e^{-ia_{\beta}(t_{1}-t_{2})} 
f(\epsilon_{r})\bar{f}(\epsilon_{l}) \nonumber \\
&\times [\mathsmaller{\sum}\limits_{\sigma=\pm}\langle a_{-}^{*}a_{+}[\varphi_{\sigma}(t_{2})-\varphi_{\sigma}(t_{1})]^{4}\rangle \nonumber \\
&-2\langle a_{-}^{*}a_{+}[\varphi_{+}(t_{2})-\varphi_{-}(t_{1})]^{4}\rangle-],
\end{align}
where $C_1 =-\lambda^{4}g_{c}/192\pi^{2}$, $g_c= \sum_n T_n$, $a_\beta = \omega_{lr}+\beta V$ and $\omega_{lr}=\epsilon_{l}-\epsilon_{r}$. 
Using Wick's theorem we write 
%where $C_1 =-\lambda^{4}g_{c}/192\hbar^{2}\pi^{2}$, $g_c= \sum_n T_n$, $a_\beta = \omega_{lr}+\beta eV/\hbar$ and $\omega_{lr}=(\epsilon_{l}-\epsilon_{r})/\hbar$. 
%Using Wick's theorem we write 
$\langle n \rangle_{\text{nq}} = 12C_{1}(n_{1}-2n_{2})$, where
\begin{align} \label{eqn1}
n_{1}&=\sum_{\sigma,\beta=\pm} \int dt_{1}dt_{2}d\epsilon_{l}d\epsilon_{r}e^{-ia_{\beta}(t_{1}-t_{2})}f(\epsilon_{r})\bar{f}(\epsilon_{l}) \nonumber \\
&\times \langle a_{-}^{*}[\varphi_{\sigma}(t_{2})-\varphi_{\sigma}(t_{1})]\rangle \langle a_{+}[\varphi_{\sigma}(t_{2})-\varphi_{\sigma}(t_{1})]\rangle \nonumber \\
&\times \langle[\varphi_{\sigma}(t_{2})-\varphi_{\sigma}(t_{1})]^{2}\rangle
\end{align}
and
\begin{align} \label{eqn2}
n_{2}&=\sum_{\beta} \int dt_{1}dt_{2}d\epsilon_{l}d\epsilon_{r}e^{-ia_{\beta}(t_{1}-t_{2})}f(\epsilon_{r})\bar{f}(\epsilon_{l}) \nonumber \\
&\times \langle a_{-}^{*}[\varphi_{+}(t_{2})-\varphi_{-}(t_{1})]\rangle \langle a_{+}[\varphi_{+}(t_{2})-\varphi_{-}(t_{1})]\rangle \nonumber \\
&\times \langle[\varphi_{+}(t_{2})-\varphi_{-}(t_{1})]^{2}\rangle.
\end{align}

A detailed derivation of $n_1$ is presented. Before computing the diagram we define the correlators in the 
frequency-domain. Using the definition 
$\varphi_{\sigma}(t)=a_{\sigma}^{*}(t)-a_{\sigma}(t)$ we have 
\begin{subequations} \label{eqA}
\begin{align}
\langle[\varphi_{\sigma}(t_{2})-\varphi_{\sigma}(t_{1})]^{2}\rangle & = -2i\int_{\omega} D_{\sigma\sigma}(\omega)[2-e^{i\omega(t_{1}-t_{2})}\nonumber \\
&-e^{-i\omega(t_{1}-t_{2})}] \\
\langle a_{-}^{*}[\varphi_{\sigma}(t_{2})-\varphi_{\sigma}(t_{1})] \rangle &=-i\int_{\omega}D_{\sigma -}(\omega)[e^{-i\omega t_{2}}-e^{-i\omega t_{1}}] \\
\langle a_{+}[\varphi_{\sigma}(t_{2})-\varphi_{\sigma}(t_{1})] \rangle &=i\int_{\omega}D_{+\sigma}(\omega)[e^{i\omega t_{2}}-e^{i\omega t_{1}}] 
\end{align}
\end{subequations}
where we used
\begin{equation}
\langle a_{+}(\omega_{1})\varphi_{\sigma}(\omega_{2})\rangle = 2\pi i D_{+\sigma}(\omega_{1})\delta(\omega_{1}+\omega_{2}),
\end{equation}
with $D_{\sigma \sigma^{\prime}}(\omega)=-i\langle a_{\sigma}(\omega) a_{\sigma^{\prime}}^{*}(\omega)\rangle$. Using Eqs.~\eqref{eqA} we rewrite Eq.~\eqref{eqn1} as
\begin{align} \label{eqS2-1}
&n_1=-\frac{2i}{(2\pi)^{3}}\sum_{\sigma,\beta}  \int dt_{1}dt_{2}d\epsilon_{l}d\epsilon_{r} d\omega d\omega_{1}d\omega_{2}
D_{\sigma\sigma}(\omega)D_{\sigma-}(\omega_{1})\nonumber \\
&\times D_{+\sigma}(\omega_{2})f(\epsilon_{r})\bar{f}(\epsilon_{l})e^{-ia_{\beta}(t_{1}-t_{2})} [e^{-i\omega_{1} t_{2}}-e^{-i\omega_{1} t_{1}}]\nonumber \\
&\times [2-e^{i\omega(t_{1}-t_{2})}-e^{-i\omega(t_{1}-t_{2})}][e^{i\omega_{2} t_{2}}-e^{i\omega_{2} t_{1}}].
\end{align}
The integrals are performed in the following sequence. First change variables $t_{1}=\tau+t_{2}$, then integrate over $t_{2}$ and $\omega_{2}$, respectively. Next, we integrate over $\tau$ obtaining $\delta$-functions that depend on $\omega$ and $\omega_{1}$. At this step we replace 
$\omega$ and $\omega_{1}$ in the $\delta$-functions by $\omega_{0}$. The smallness of $\kappa_0$ in comparison with $\omega_0$ validates this approximation. Finally, we integrate over the energies and use the definition
\begin{align} \label{nosdef}
\sum_{\beta}&\int d\epsilon_{l} d\epsilon_{r} f(\epsilon_{r})\bar{f}(\epsilon_{l}) \delta(a_{\beta}-\omega) =\bar{S}(\omega)
%\sum_{\beta}&\int d\epsilon_{l} d\epsilon_{r} f(\epsilon_{r})\bar{f}(\epsilon_{l}) \delta(a_{\beta}-\omega) =\hbar \bar{S}(\omega)
\end{align}
Equation~\eqref{eqS2-1} is now written as
\begin{align} \label{eqS2-3}
&n_{1}=-\frac{i}{\pi}[6\bar{S}(0)-4\bar{Y}(\omega_{0})+\bar{Y}(2\omega_{0})]  \nonumber  \\
&\times \sum_{\sigma} \int d\omega d\omega_{1} D_{\sigma\sigma}(\omega)D_{\sigma-}(\omega_{1})D_{+\sigma}(\omega_{1}), 
\end{align}
where we defined $\bar{Y}(\omega_{0})=\bar{S}(\omega_{0})+\bar{S}(-\omega_{0})$. The final step is to perform the integrals over $\omega$ and $\omega_{1}$. 
With the help of the relation $a_\pm = (a_{c}\pm a_q)/\sqrt{2}$ we rewrite the GFs $D_{\sigma \sigma^\prime}$ in terms of the Keldysh, retarded 
and advanced GFs defined in Sec.~\ref{stheory} considering $T_n \ll 1$. The result of these integrals is
\begin{subequations} \label{intgs}
\begin{align}
&\int d\omega D_{\sigma \sigma}(\omega) = -\frac{i\pi}{2\omega_0}\bar{Y}(\omega_0) \\
&\sum_{\sigma}\int d\omega_{1} D_{\sigma-}(\omega_{1})D_{+\sigma}(\omega_{1})=-\frac{\pi^{2}\bar{S}(-\omega_{0})}{\lambda^{2}g_{c}\omega_{0}^{3}}\bar{Y}(\omega_0)
\end{align}
\end{subequations}
Replacing Eqs.~\eqref{intgs} into Eq.~\eqref{eqS2-3} we obtain 
\begin{equation} \label{eqS2-5}
n_{1}=\pi^{2}\frac{\bar{Y}^{2}(\omega_{0})}{2\lambda^{2}g_{c}\omega_{0}^{4}}\bar{S}(-\omega_{0})[6\bar{S}(0)-4\bar{Y}(\omega_{0})+\bar{Y}(2\omega_{0})].
\end{equation}

Following the same steps presented above we compute $n_2$, resulting in 
\begin{align}
&n_{2}= \frac{\pi^{2}}{2\lambda^2 g_{c} \omega_{0}^{4}}\{2\bar{S}(-\omega_0)\bar{Y}(\omega_0)[\bar{Y}(\omega_{0})\bar{S}(0)  \\
&-2\bar{S}(\omega_{0})\bar{S}(-\omega_{0})]-2\bar{S}^{2}(-\omega_0)[\bar{Y}(\omega_{0})\bar{S}(\omega_{0})\nonumber \\
&-\bar{S}(-\omega_{0})\bar{S}(2\omega_{0})-\bar{S}(\omega_{0})\bar{S}(0)] -[\bar{S}^{2}(-\omega_0)+\bar{S}^{2}(\omega_0)] \nonumber \\
&\times [\bar{Y}(\omega_{0})\bar{S}(-\omega_{0})-\bar{S}(-\omega_{0})\bar{S}(0)-\bar{S}(\omega_{0})\bar{S}(-2\omega_{0})]\}. \nonumber
\end{align}
After some algebraic manipulation we show that
\begin{align}
n_1-2n_2 &= \frac{2\pi^{2}}{\lambda^{2}g_{c}\omega_{0}^{3}}[\bar{S}^{2}(-\omega_0)\bar{S}(2\omega_0)-\bar{S}^{2}(\omega_0)\bar{S}(-2\omega_0)].
\end{align}
Finally, $\langle n \rangle_{\text{nq}} = 12C_{1}(n_{1}-2n_{2})$ gives the non-quadratic contribution to mean number of photons Eq. \eqref{npng}.

\section{Transition rate \label{appC}}

In this section we derive the coefficients $\Gamma_{i\rightarrow j}$ of the rate equation model presented in Sec.~\ref{sec3}. They are obtained via Fermi golden rule
\begin{equation} \label{eqtrate}
\Gamma_{i\rightarrow j} = \frac{2\pi}{\hbar}|\langle f | \bar{H}_\text{T}| i \rangle |^2\delta(E_{i}-E_{f}).
\end{equation}
This equation describes the transition from the state $|i\rangle$ to $|f\rangle$. $E_{i(f)}$ is the energy of the state $|i\rangle$ ($|f\rangle$), and 
the interacting Hamiltonian is
\begin{equation} \label{tun1}
\bar{H}_{T}(\lambda)=\hat{\mathcal{T}}e^{-\lambda(\hat{a}^{\dagger}-\hat{a})}+\hat{\mathcal{T}}^{\dagger}e^{\lambda(\hat{a}^{\dagger}-\hat{a})}.
\end{equation}
As we are considering QPC with tunneling probabilities $T_n\ll 1$, the tunneling operator is well describe by 
$\hat{\mathcal{T}}=\gamma \sum_\mathbf{k,q} c_{l,\mathbf{k}}^{\dagger}c_{r,\mathbf{q}}$,
where $\gamma$ is the tunneling amplitude and $c_{\alpha,\mathbf{k}}^{\dagger}$ is the electron creation operator in the reservoir $\alpha$. 
Considering $\lambda \ll 1$, we expand the exponential to second-order and rewrite Eq.~\eqref{tun1} as
\begin{align} \label{tunexp}
\bar{H}_\text{T} &= \left(1-\frac{\lambda^{2}}{2}(2\hat{n}+1)\right)\bar{H}_{T}(0) +\lambda(a^\dagger -a)\hat{I} \nonumber \\
&+ \frac{\lambda^2}{2}(a^{\dagger 2} +a^{2})\bar{H}_{T}(0),
\end{align}
with $\hat{n}$ the photon number operator, and $\hat{I}=\gamma \sum_\mathbf{k,q} (c_{r,\mathbf{q}}^{\dagger}c_{l,\mathbf{k}}-\text{h.c})$. 
The first term does not produce any transition, and hence, it is neglected. The second and third terms give rise to transitions from the state 
$|n\rangle$ to $|n\pm1\rangle$ and $|n\pm 2\rangle$ states, respectively. As expected, the expansion in the e-p coupling shows that the 
transitions to the state $|n\pm 1\rangle$ and $|n\pm 2\rangle$ will be proportional to $\lambda^{2}$ and $\lambda^{4}$, respectively.

Considering $|i\rangle = |n \rangle \otimes |E_{g}\rangle$ and $|f\rangle = |m \rangle \otimes |E_{e}\rangle$, in which $|n \rangle$ is the 
Fock state of $n$-photons with energy $E_n=n\hbar\omega_{0}$. $|E_{g}\rangle$ and $|E_{e}\rangle$ are the electron states with respective 
energies $E_{g}$ and $E_{e}$. 

The coefficient $\Gamma_{n\rightarrow n\pm 1}$ is obtained replacing $\bar{H}_\text{T}$ in Eq.~\eqref{eqtrate} by the second term of 
Eq.~\eqref{tunexp}. It is equal to
\begin{align} \label{trans1}
&\Gamma_{n \rightarrow n\pm 1} = \frac{2\pi\lambda^{2}}{\hbar}|\langle n\pm 1 | (a^{\dagger} -a)| n \rangle |^2 |\langle E_e | \hat{I}| E_g \rangle |^2 \nonumber \\
&\times \delta(E_{g}-E_{e} \mp \hbar \omega_0) \nonumber \\
&=\frac{2\pi\lambda^{2}}{\hbar}\left(n+\frac{1}{2}\pm\frac{1}{2}\right)|\langle E_e | \hat{I}| E_g \rangle |^2 \delta(E_{g}-E_{e} \mp \hbar \omega_0). 
\end{align}
We need to evaluate the matrix element $\langle E_e | \hat{I}| E_g \rangle$. For that let us consider the matrix element 
$\langle E_e | c_{r,\mathbf{q}}^{\dagger}c_{l,\mathbf{k}}| E_g \rangle$. It is different of zero only if
\begin{subequations} 
\begin{align}
| E_g \rangle & = |0,\ldots, 0_{r,\mathbf{q}}, \ldots, 1_{l,\mathbf{k}}, \ldots \rangle \\
| E_e \rangle & = |0,\ldots, 1_{r,\mathbf{q}}, \ldots, 0_{l,\mathbf{k}}, \ldots \rangle,
\end{align}
\end{subequations} 
i.e, the initial state $| E_g \rangle$ the state $\mathbf{k}$ in the left lead must be occupied, while the state $\mathbf{q}$ in 
the right lead must be empty. This implies that $|\langle E_e | c_{r,\mathbf{q}}^{\dagger}c_{l,\mathbf{k}}| E_g \rangle|^2 = f(\epsilon_k)\bar{f}(\epsilon_q)$ with $E_g=\epsilon_k$, $E_s = \epsilon_q$, $f(\epsilon_k)$ the Fermi function and $\bar{f}(\epsilon_q)=1-f(\epsilon_q)$. A similar expression is found for second term of the $\hat{I}$. Therefore, we write 
\begin{align} \label{ele-matrix}
&|\langle E_e | \hat{I}| E_g \rangle|^2\delta(E_{g}-E_{e} \mp \hbar \omega_0)= \rho_0^2 \gamma^2 \int d\epsilon_k d\epsilon_q [  f(\epsilon_k)\bar{f}(\epsilon_q) \nonumber \\
&\times \delta(\epsilon_k-\epsilon_q \mp \hbar \omega_0 +V) + f(\epsilon_q)\bar{f}(\epsilon_k)\delta(\epsilon_k-\epsilon_q \mp \hbar \omega_0 -V)]\nonumber \\
&=\rho_0^2\gamma^2 \bar{S}(\mp\omega_0)
\end{align}
where $\rho_{0}$ is the electronic density of the state, which we assume to be the same for both leads. We considered that the chemical potential 
in the left (right) reservoir is $\mu_l = V/2$ ($\mu_r = -V/2$). Thus, to obtain Eq.~\eqref{ele-matrix} we shift $\epsilon_{k(q)} \rightarrow \epsilon_{k(q)} \pm V/2$. The integrals over the energies were preformed with the help of Eq.~\eqref{nosdef}. Therefore, the transition 
rate from the Fock state $|n\rangle$ to $|n\pm 1\rangle$, Eq.~\eqref{trans1}, is
\begin{equation} \label{ematrxi1}
\Gamma_{n \rightarrow n\pm 1} =\frac{2\pi\lambda^{2}\rho_0^2\gamma^2}{\hbar}\left(n+\frac{1}{2}\pm\frac{1}{2}\right)\bar{S}(\mp\omega_0)
\end{equation}
As expected the transition rate from $|n\rangle$ to $|n+1\rangle$ ($|n-1\rangle$) is proportional to the QPC emission (absorption) noise 
power of one-photon.

Analogously, we obtained the transition coefficients $\Gamma_{n\rightarrow n\pm 2}$, which are 
\begin{equation} \label{ematrxi2}
\Gamma_{n \rightarrow n\pm 2} =\frac{\pi\lambda^{4}\rho_0^2\gamma^2}{2\hbar}\left(n+\frac{1}{2}\pm\frac{1}{2}\right)\left(n+\frac{1}{2}\pm\frac{3}{2}\right)\bar{S}(\mp2\omega_0).
\end{equation}
The transition rate from $|n\rangle$ to $|n+2\rangle$ ($|n-2\rangle$) is characterized the QPC emission (absorption) of two-photon. 
These transitions coefficients are next-to-leading order correction to the cavity absorption and emissions processes. 

Finally, the rate equation, Eq.~\eqref{rateq0}, is fully characterized by the transitions rates defined in Eqs.~\eqref{ematrxi1} and \eqref{ematrxi2}.

%%%\bibliography{biblio}

%

\end{document}